\documentclass[backref, breaklinks, twocolumn, colorlinks]{aastex631}   
\hypersetup{linkcolor=blue,citecolor=blue,filecolor=cyan,urlcolor=black}

\usepackage{booktabs}
\usepackage{gensymb}

\usepackage{latexsym,amsmath,amssymb}
\usepackage{verbatim}
\usepackage{multirow}
\usepackage{color}
\usepackage[hang]{footmisc}
\usepackage{soul,xcolor}
\usepackage{afterpage}
\usepackage[counterclockwise]{rotating}
   \usepackage{amssymb}
   \usepackage{amsmath}

          \font\sixrm=cmr6

\newcommand {\xmm} {\textsl{XMM-Newton}}

\newcommand {\nustar} {\textsl{NuSTAR}}

\newcommand {\nicer} {\textsl{NICER}}
\newcommand {\ixpe} {\textsl{IXPE}}

\def\teq#1{$\, #1\,$}
\def\rns{R_{\hbox{\sixrm NS}}}
\def\mns{M_{\hbox{\sixrm NS}}}

\def \src {1RXS~J1708$-$40}

\begin{document}

\title{Detailed Timing, Spectral, and Polarimetric Analysis of Magnetar 1RXS J170849.0-400910}

\author[0000-0002-0254-5915]{Rachael Stewart}
\affiliation{George Washington University, 725 21st St NW, Washington, DC 20052, USA}
\correspondingauthor{Rachael Stewart}
\email{raestewart@gwu.edu}

\author[0000-0002-7991-028X]{George~Younes}
\affiliation{Center for Space Sciences and Technology, University of Maryland, Baltimore County, Baltimore, MD 21250}
\affiliation{Astrophysics Science Division, NASA Goddard Space Flight Center, Greenbelt, MD 20771, USA}

\author[0000-0001-6119-859X]{Alice K. Harding}
\affiliation{Theoretical Division, Los Alamos National Laboratory, Los Alamos, NM 87545, USA}

\author[0000-0001-9268-5577]{{Hoa} {Dinh Thi}}
\affiliation{Department of Physics and Astronomy—MS 108, Rice University, 6100 Main St., Houston, TX 77251-1892, USA}

\author[0000-0003-4433-1365]{{Matthew G.} {Baring}}
\affiliation{Department of Physics and Astronomy—MS 108, Rice University, 6100 Main St., Houston, TX 77251-1892, USA}

\author[0000-0002-9249-0515]{{Zorawar} {Wadiasingh}}
\affiliation{Department of Astronomy, University of Maryland, College Park, MD 20742,USA}
\affiliation{Astrophysics Science Division, NASA Goddard Space Flight Center, Greenbelt, MD 20771, USA}

\author[0000-0002-6548-5622]{{Michela} {Negro}}
\affiliation{Department of Physics and Astronomy, Louisiana State University, 202 Nicholson Hall, Baton Rouge, LA, 70803, USA}

\author[0000-0002-3905-4853] {{Alex} {Van Kooten}}\affiliation{George Washington University, 725 21st St NW, Washington, DC 20052, USA}



\begin{abstract}
We present a broadband timing, spectral, and polarimetric study of the magnetar 1RXS~J170849.0$-$400910 using \xmm, \nustar, and \ixpe. The pulse morphology evolves strongly across 0.5--70 keV. Below 3 keV, the emission is dominated by a broad soft pulse with a leading shoulder that develops into a faint interpulse near 3 keV, while the pulse fraction remains  $\approx25$\%. The profile becomes increasingly double-peaked between 3 and 20 keV and returns to a single peak at higher energies. The pulse fraction dips to $\sim20\%$ near 4 keV and rises to $\sim42\%$ above 25 keV. The phase-averaged spectrum is well described by an absorbed blackbody plus two power-laws, with $kT=0.468\pm0.003$ keV, $\Gamma_{\rm soft}=2.63\pm0.04$, and $\Gamma_{\rm hard}=0.5\pm0.1$. Phase-resolved spectroscopy reveals distinct soft and hard pulse components. The thermal modulation is driven primarily by a factor of $\sim5$ variation in projected emitting area, whereas the soft power-law exhibits two peaks with different phase and energy evolution, suggesting distinct emission regions or mechanisms. The 10--70 keV flux is strongly anticorrelated with the soft power-law photon index, linking spectral hardening to the hard pulse. The polarization degree also varies strongly with phase and energy. In the 2--3 keV band, it is anticorrelated with the intensity profile, consistent with magnetized-atmosphere emission, whereas in the 4--8 keV band it reaches $64\pm10\%$ during the nonthermal power-law-dominated peak. This high polarization can be reproduced by magnetospheric quantum pair-synchrotron emission. Together, these results reveal an intricate, phase-dependent superposition of emitting regions and radiative processes whose complexity emerges only through broadband, phase-resolved spectropolarimetry.
\end{abstract}



\section{Introduction} \label{sec:intro}
Magnetars are a subclass of isolated neutron stars which are typically characterized by their slow periods ($P \sim 1-12$ s), large period derivatives ($\dot{P} \sim 10^{-13}-10^{-11} $s s$^{-1}$), and high X-ray luminosities ($L_X \sim 10^{31}-10^{36}$ erg s$^{-1}$). The observed rotational characteristics suggest that magnetars possess ultra-powerful dipolar magnetic fields inferred to be $\sim 10^{14}-10^{15}$ G (\citealt{1992ApJ...392L...9D}; \citealt{1993ApJ...408..194T}; \citealt{1998Natur.393..235K}). The decay of these super-critical magnetic fields serve as the predominant source powering the magnetars' soft and hard X-ray emission (see \citealt{2015RPPh...78k6901T}, \citealt{2017ARA&A..55..261K}, for recent reviews).

The persistent (pulsed) high-energy emission of magnetars can be characterized by two spectral peaks. The first occurs in the soft X-ray band (0.5--10 keV) where thermal photons emitted from the hot surface are modified by a dense highly-magnetized atmosphere. The second occurs in the hard X-ray regime ($\gtrsim10$ keV) where distinctive nonthermal tails extend beyond 100 keV, suggesting that some magnetars' energy output is predominantly emitted in the hard X-rays to soft $\gamma$-rays \citep[e.g.,][]{2008A&A...489..263D}.

The X-ray radiation emanating from magnetars is expected to be highly polarized (\citealt{Heyl-2000-MNRAS}, \citeyear{Heyl-2002-PRD}). Linearly polarized photons adopt either the ordinary (O) mode, parallel to the plane of the external magnetic field and photon wave vector, or the extraordinary mode (X), perpendicular to the same plane. High energy photons propagating through the tenuous highly magnetized plasma surrounding the magnetar will retain their incident polarization state due to the QED effect, vacuum birefringence \citep{2006RPPh...69.2631H}. The polarization degree (PD) of the emergent radiation is therefore dependent upon various elements including the atmospheric structure, surface temperature profile, and properties of the magnetosphere. Thus spectro-polarization analysis enables insights into geometry of the magnetar system and the radiative mechanisms driving the empirical phenomenology observed in their spectra \citep[see, e.g.,][]{taverna24Galax}.  

The magnetar 1RXS~J170849.0$-$400910 (hereafter \src) exhibits one of the brightest persistent X-ray fluxes ($F_{\mathrm{1-10 keV, abs}}\sim 4 \times 10^{-11}$ erg s$^{-1}$ cm$^{-2}$) observed among the population. It was first detected in the \textit{ROSAT} All-Sky Survey in 1996 with follow-up observations performed by \textit{Advanced Satellite for Cosmology and Astrophysics} (\textit{ASCA}) in 1997 (\citealt{1996rftu.proc..637V}, \citealt{1997IAUC.6585....2S}). It has been observed as part of a long term monitoring program ever since (\citealt{2014ApJ...784...37D}). \src\ possesses a period of $\sim 11$ s, $\dot{P}$ of $1.9 \times 10^{-11} $s s$^{-1}$, characteristic age of $9$ kyr, and dipolar magnetic field strength of $\sim 10^{14}$ G (\citealt{1999ApJ...518L.107I}; \citealt{2002ApJ...567.1067G}). 1RXS J1708--40 also exhibits bright pulsed emission at hard X-rays, first detected by the \textit{INTErnational Gamma-Ray Astrophysics Laboratory} (\textit{INTEGRAL}; \citealt{2004AstL...30..382R}; \citealt{2006ApJ...645..556K}).  It additionally displays a high polarization degree of $\sim 35\%$ between 2--8 keV \citep{2023ApJ...944L..27Z}. Thus, 1RXS J1708--40 is an excellent candidate for targeted broadband X-ray spectroscopic and spectro-polarimetric observations.

In this paper, we present the detailed persistent emission characteristics of the magnetar 1RXS J1708--40 using the \textit{Nuclear Spectroscopic Telescope Array} (\nustar), the \textit{X-ray Multi-Mirror Mission} (\xmm) telescope, and the \textit{Imaging
X-ray Polarimetry Explorer} (\ixpe). 
Section \ref{section2} describes the observations and the data reduction process. In Section \ref{section3} we report the temporal characteristics along with the phase-averaged and phase-resolved spectroscopic and polarimetric results. Section \ref{section4} summarizes the main findings of our study and provides a comparison with previous analyses. The former are then discussed in the context of the magnetar theoretical expectations in Section~\ref{sec:discussion}.

\begin{table*}[t!]
\label{tab:J1708ObsSummary}

\begin{center}
\resizebox{1.0\textwidth}{!}{
\hspace*{-1.75cm}
\begin{tabular}{l c c c| c c c}

\midrule
\hline
Observation ID & Telescope & Date (UTC)& Livetime Exposure & $\nu$ & $\dot{\nu}$ & $\mathrm{P_{epoch}}$\\

 & & & (ks) & (Hz) & ($10^{-13}$~Hz~s$^{-1}$) & (MJD)\\
\midrule

\hline
30401023002/0830400101 & \textit{NuSTAR/XMM} & 2018-08-30 00:34:39.880 & 93.274/36.391 & 0.090817546 & -- & 58360.0240727 \\
90502353002& \nustar\ & 2019-12-09 04:00:15.900 & 21.478 & 0.090810948 & -- & 58826.1668507 \\
90502353004& \nustar\ & 2019-12-30 01:09:44.654 &24.934 & 0.090810643 & -- & 58847.0484335 \\
01003199 & \ixpe\ & 2022-09-28 00:00:01.668 & 837 & 0.090795763 & -1.9523 & 59850.0000193 \\
\hline
\end{tabular}}
\caption{All timing solution frequencies and epochs above are in Barycentric Dynamical Time (TDB). For the \nustar\ timing solutions, adding a second frequency derivative term did not significantly improve the fit so they were omitted during the folding.}
\label{table:timing_solutions}
\end{center}
\end{table*}

\section{Observations and Data Reduction}{\label{section2}} 

\src\ was observed simultaneously with \nustar\ (ObsID 30401023002) and \xmm\ (ObsID 0830400101) starting on August 28, 2018 for approximately 93.3 ks and 36.4 ks of total livetime exposure, respectively. To improve the signal-to-noise ratio (S/N) of the source in the hard X-rays we also analyze two \nustar\ Target of Opportunity (ToO) observations taken on December 8, 2019 (ObsID 90502353002) and December 29, 2019 (ObsID 90502353004). 1RXS J1708--40 was also observed as a primary science target for \ixpe\ on September 19, 2022 (ObsID 01003199) for 837 ks of livetime exposure. We provide a summary of the above observations in Table \ref{tab:J1708ObsSummary}.
 
\subsection{NuSTAR}
\nustar, a focusing X-ray telescope sensitive to 3--79 keV, is comprised of two identical focal plane modules, FPMA and FPMB \citep{2013ApJ...770..103H}. We reduced the \nustar\ data using \texttt{nustardas} v2.1.2 with CALDB files version 20220525. The 2018 observation contained a flare that was filtered out, enabling focused analysis of the source's persistent emission (for details of the flare analysis, see \citealt{2020ApJ...889L..27Y}). SAA filtering was performed using the \texttt{nustardas} task \texttt{nustarsaa}  with the flags \texttt{saamode=optimized}, \texttt{tentacle = no}, \texttt{saacalc=1} to correct for increased background activity in the GTIs related to passage through the South Atlantic Anomaly. 

We employed the nustar-gen-utils module \texttt{nustar\_gen.radial\_profile} to determine that a circular region of 60'' maximizes the S/N in this observation. For estimation of the background, we selected a source-free circular region, of the same size from the same detector as the source. 

High level data-products, i.e., light curves, spectra, response matrix files, and ancillary files, were extracted using \texttt{nuproducts}. The task \texttt{nuproducts} performed additional cleaning and calibration including barycenter-correcting the photon time of arrivals, \nustar\ clock drift corrections, and livetime, point-spread function (PSF), and vignetting corrections to the light curves. 

We applied the same approach outlined above to process the two ToO observations (90502353002 and 90502353004) apart from two key differences. 
Firstly, a glitch occurred between the \nustar+\xmm\ observation and the ToO observations. We thus investigated the ToO spectral properties to determine whether there was any notable radiative variability by simultaneously fitting the ToO phase-averaged spectra with the \nustar+\xmm\ spectra. This is further detailed in Section \ref{sec:specanalysis}. 
Secondly, the two ToO observations were taken when the source was close to the Sun (e.g. at angles $<45\degree$), causing the on-board star tracker to be blinded. Thus, the \texttt{nustardas} task \texttt{nusplitsc} was required to generate cleaned event files for each Camera Head Unit (CHU) combination\footnote{For additional information about \texttt{nusplitsc}, see \url{https://heasarc.gsfc.nasa.gov/docs/software/lheasoft/help/nusplitsc.html.}}.
After filtering and cleaning the two ToOs, they contribute a combined additional 13 ks to the total livetime exposure. 

\subsection{XMM-Newton}
\xmm\ is a soft X-ray observatory operating between 0.3 and 12 keV \citep{2001A&A...365L...1J}. It observed \src\ (ObsID 0830400101) simultaneously with \nustar\ (ObsID 30401023002). The EPIC-pn telescope operated in Small Window mode with the medium filter. We did not include the MOS telescope data in favor of the higher timing resolution of the EPIC-pn cameras. The EPIC-pn data was reduced using \xmm\ Science Analysis System (SAS) version 21.0 with  calibration files (CCF) from April 2023 and HEASOFT version 6.31. 

Data were filtered according to PATTERN 0--4 from good events (``FLAG=0").  We excluded high particle background levels, e.g. background rates above 10 keV exceeding $0.4$ ct/s. The task \texttt{eregionanalyse} calculated the source centroid and extraction region size that optimizes the source S/N. For the source, we used a circular 60'' region, and for the background region we used a source-free annulus with an inner and outer radii of 175" and 250", respectively. 

We extracted light curves and spectra in the 0.5–10 keV energy range using \texttt{evselect}. The XMM-SAS tasks \texttt{rmfgen} and \texttt{arfgen} produced the response matrix files and ancillary response files, respectively. We applied \texttt{epiclccorr} to correct the light curves for PSF variations, vignetting, and bad pixels. 

\subsection{IXPE}
\ixpe, the first X-ray polarimetry space observatory, consists of three identical detector units (DUs) observing in the 2--8 keV energy band \citep{2022JATIS...8b6002W}.  We retrieved the archival observation of \src\ (ObsID 01003199). We then used ds9 to select a 60'' circular region for the source and an annulus of 120'' and 240'' for the inner and outer radii respectively for the background region. 

For all other cleaning, data product extraction, and analysis we used the software \texttt{ixpeobssim} \citep{baldini_2022_softX_ixpe}. We set the instrument response functions (IRFs) to \texttt{ixpe:obssim20220702:v13} from \texttt{ixpeobssim}'s pseudo-CALDB which contains the arf, rmf, PSF, vignetting functions, modulation factors, and modulation responses corresponding to the observation date \citep{2022SoftX..1901194B}. Folded event files were produced using \texttt{xpphase}. The task \texttt{xpbin} extracted all binned products, including the pulse profiles and the polarization cubes. Lastly, \texttt{xpselect} created the energy-resolved data products, e.g., the event files and pulse profiles in predefined energy bands. 

\section{Results}{\label{section3}}

\subsection{Timing Analysis} \label{sec:floats}

We barycentered all time-tagged events to the solar system barycenter using the \src\ localization as reported in \citet{israel03ApJ:1708} and the JPL ephemerides DE405. We phase folded the events using a phase-connected timing solution derived from a long-term \nicer\ monitoring, which extend over all of our observation epochs (van Kooten et al. in prep.). Table \ref{table:timing_solutions} summarizes the spin ephemerides that are valid across each observation as derived from the full timing solution. The phase $\phi_0$ of the pulse profiles was arbitrarily chosen to be where the \xmm\ 0.5--10 keV pulse profile reached a maximum.   

\begin{figure}[t!]
    \centering
    \makebox[0pt]
    {\includegraphics[width=1.\linewidth]{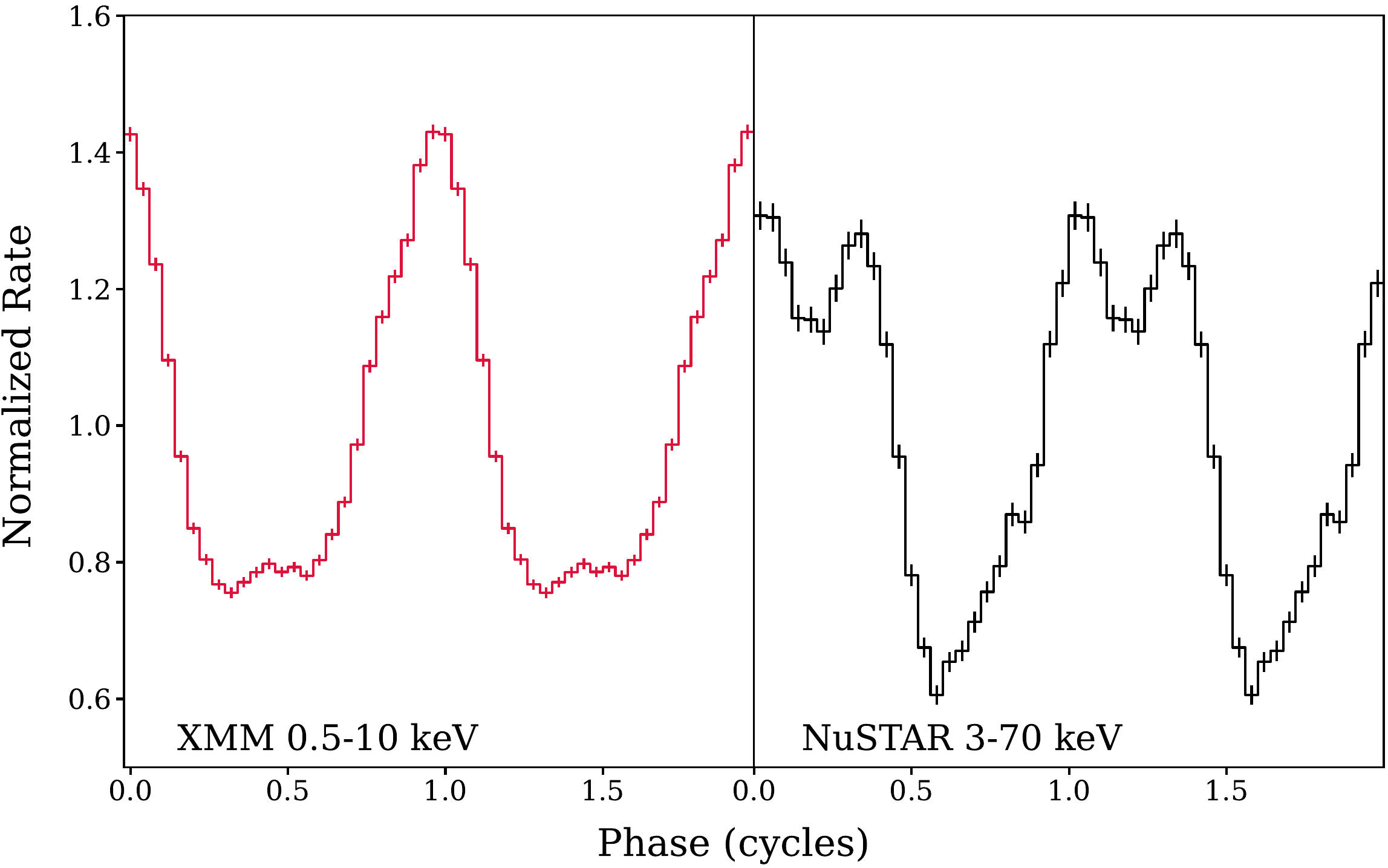}}
    \caption{\xmm\ (red) and \nustar\ FPMA + FPMB (black) energy-integrated and source-filtered pulse profiles of \src. Two rotational cycles are displayed for clarity.}
    \label{fig:energy_integrated_pp}
\end{figure}

\begin{figure*}[t!]
    \centering
    \makebox[0pt]
    {\includegraphics[width=1.1\linewidth]{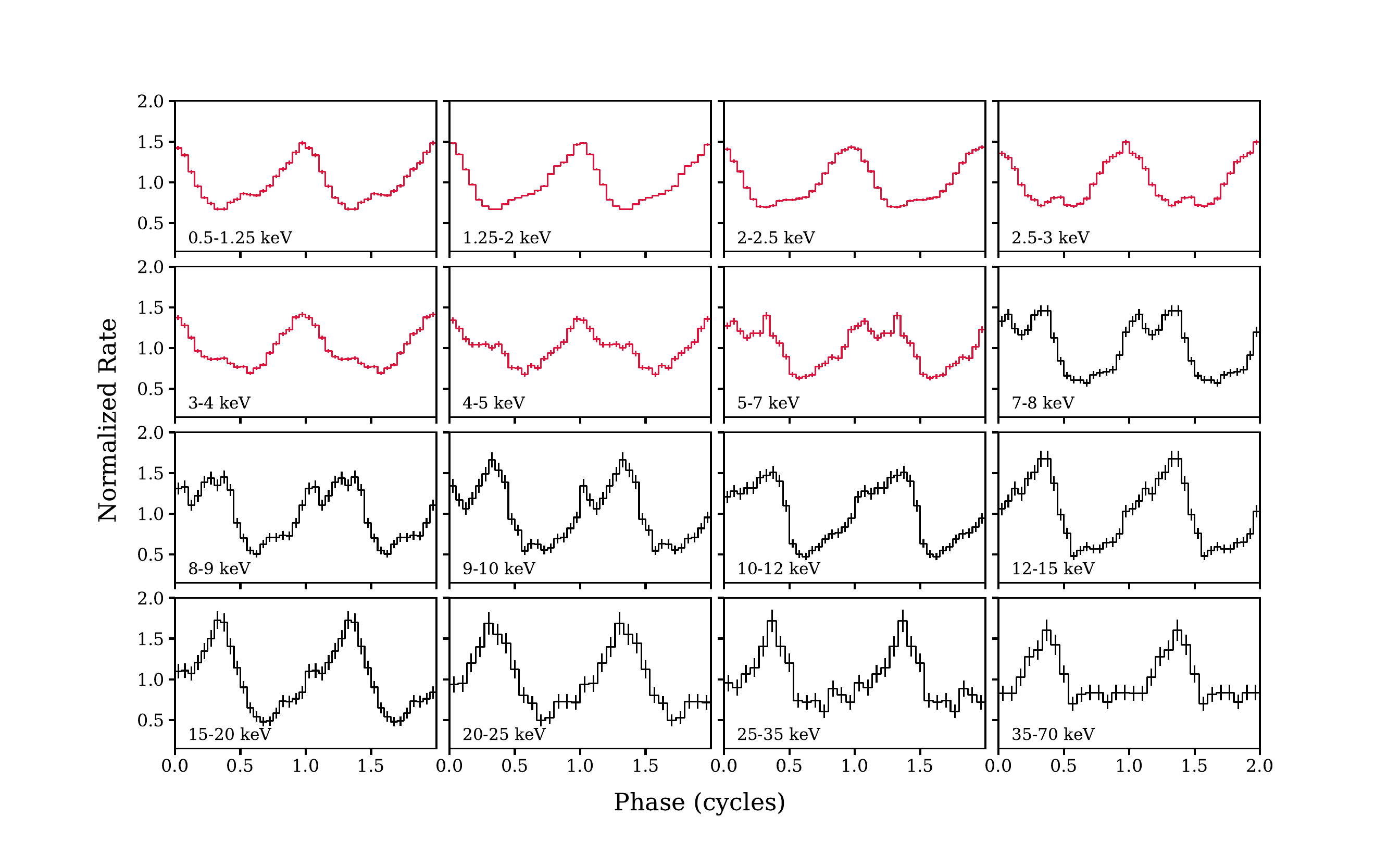}}
    \caption{Energy-dependent pulse profiles of \src. \xmm\ are shown in red and \nustar\ FPMA + FPMB are shown in black. All pulse profiles are folded into 20 bins apart from the last three (20--25, 25--35, and 35--70 keV) which are binned to 15 phase bins.}
    \label{fig:nuxmm_energy_dependent_pp}
\end{figure*}

Figure \ref{fig:energy_integrated_pp} shows the energy-integrated \xmm\ (0.5--10 keV) and \nustar\ (3--70 keV) folded pulse profiles of 1RXS J1708--40. Both profiles are normalized by their average. Note the presence of a secondary peak in the \nustar\ profile coincident with the pulse minimum in the \xmm\ one, near $\phi\approx0.3$. We divided the \nustar+\xmm\ pulse profiles into 16 energy bands, as shown in Figure \ref{fig:nuxmm_energy_dependent_pp}. Each profile uses 20 phase bins except for the 20--25, 25--35, and 35--70 keV bands which are binned according to 15 phase bins due to fewer counts at higher energy ranges. The pulse morphology exhibits large variability across the energy bands. At the softest energies from 0.5~keV to 1.25~keV, the profile is dominated by a single peak. Yet, there is a prominent shoulder in the leading edge of the main peak. This shoulder transitions into a secondary peak as observed in the energy band 2.5--3~keV. Notably, these two peaks are separated by approximately 0.5 cycles. 

At energies $>3$ keV, a shoulder starts appearing in the trailing edge of the main pulse. This shoulder increases in prominence with energy and emerges as its own peak for energies $>5$~keV, at phase $\phi\approx0.3$. Moreover, the initial primary peak seen at the softest energies appears to slightly shift towards the right at higher energies when comparing the first three rows in each column of Figure~\ref{fig:nuxmm_energy_dependent_pp} (see also Section~\ref{sec:prs}), while also decreasing in power. The profile becomes single-peaked at energies $>20$~keV, while the duty cycle decreases; in the energy range 35--70 ~keV, we measure a duty cycle of about 45\%.


We measured the root-mean-squared pulsed fraction (RMS PF) as described in, e.g., \citet{2014ApJ...784...37D}. For this, we first produced light curves in a given energy band binned at 16 ms corrected with \texttt{epiclcorr} and \texttt{nulccorr} for the \xmm\ and \nustar\ data, respectively (see Section~\ref{section2}). We then assigned phases to each time bin according to the timing solution referenced in Table \ref{table:timing_solutions} and folded them to create a pulse profile with 20 phase bins. We subsequently subtracted the average background from these pulse profiles as derived in the same energy band. Finally, we included the contribution from the first five harmonics in our RMS PF calculation. The 0.5--10~keV \xmm\ and 3--70~keV \nustar\ PFs are $(0.25 \pm 0.03)$ and $(0.34 \pm 0.04)$, respectively.

\begin{figure}[ht!]
    \includegraphics[width=\linewidth]{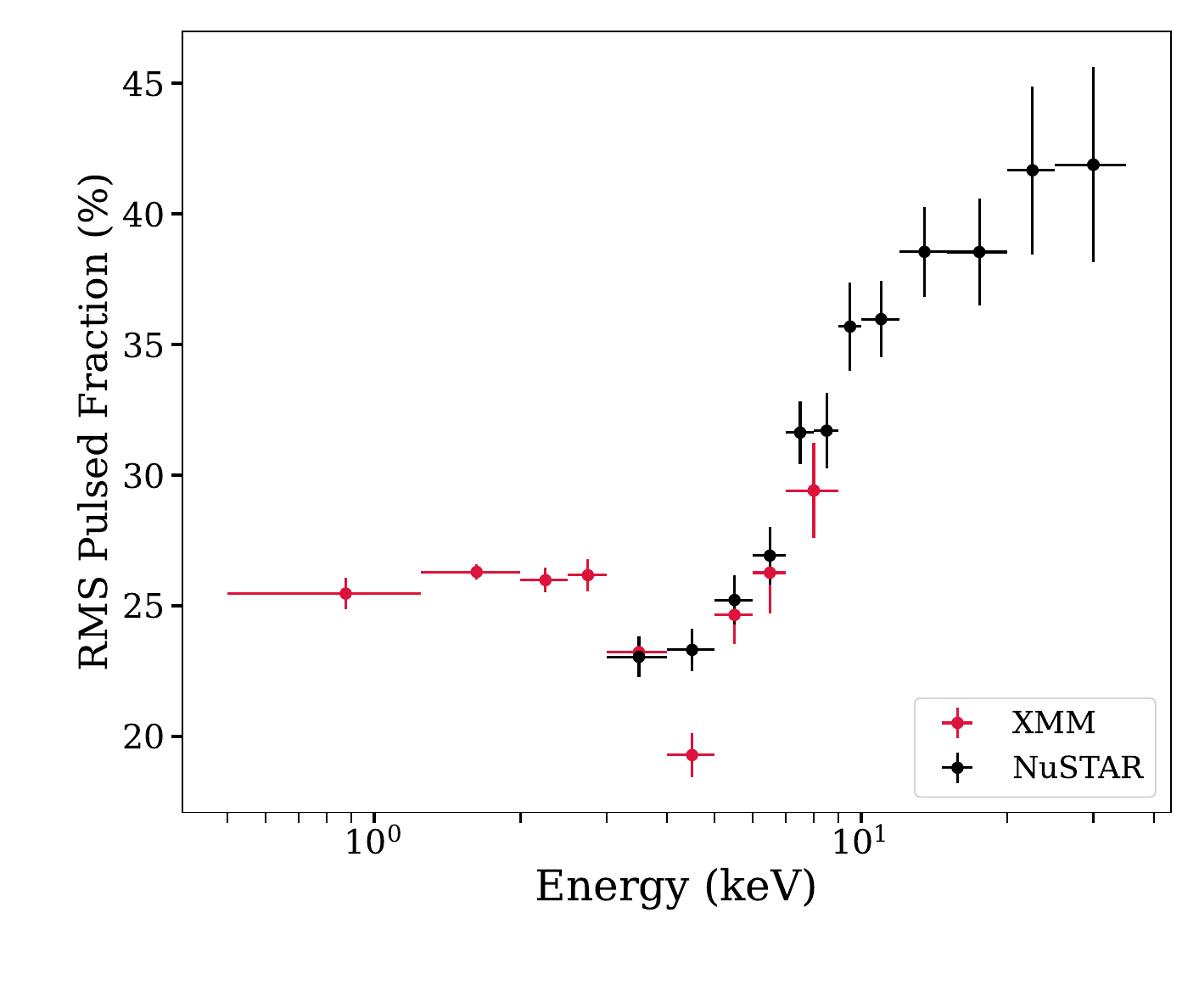}
    \caption{Energy dependent, background-subtracted RMS PF of simultaneous \xmm\ (red) and \nustar\ (black) observation of \src.}
    \label{fig:rms_pf_energy_resolved}
\end{figure}

We similarly extracted the RMS PFs in fine energy bands for \xmm\ and \nustar\ as shown in Figure~\ref{fig:rms_pf_energy_resolved}. We find that the PF remains largely stable from 0.5 keV to 3 keV at $\approx 25\%$. It abruptly decreases at energies $>3$~keV, and reaches a minimum of $\approx20\%$ at around 4~keV. The PF starts linearly increasing from about $5$~keV to 12~keV, then begins to flatten settling at around $42\%$ at energies $>25$~keV\footnote{Note that the \xmm\ and \nustar\ RMS PF in the same energy bands are consistent at the 1$\sigma$ level, except in the energy band 4--5~keV. Given that our error bars are statistical only, we attribute the difference to unaccounted-for systematic uncertainties between the two instruments.}.
 

\subsection{Spectral Analysis}\label{sec:specanalysis}

Our spectral analyses were performed using Xspec version 12.13.0c as part of HEASOFT version 6.31 \citep{1996ASPC..101...17A}. To account for absorption of X-ray photons in the interstellar medium (ISM), we utilize the Wilms abundances \citep{2000ApJ...542..914W}, the photo-electric cross-sections of \citealt{1996ApJ...465..487V}, and the Tuebingen-Boulder ISM absorption model (\texttt{tbabs}). For each \nustar\ and \xmm\ observation, we grouped all spectra to have at least 5 counts per bin, and fit the spectral models using the W-statistic (contained within \texttt{cstat} in Xspec) and the Anderson-Darling (\texttt{ad}) test statistic. For all our spectral fits, we introduced a cross-normalization constant to track calibration uncertainties between the different telescopes. We find the uncertainties to be approximately 4\% between the two \nustar\ instruments and between 10\% and 20\% between \nustar, \xmm, and \ixpe\ which aligns with the expected cross-instrument calibration uncertainties \citep{madsen_2022_iachec}. Finally, in the following we report all uncertainties at the 1$\sigma$ level unless otherwise stated.

A spin-up glitch occurred between the simultaneous \nustar+\xmm\ observation and the \nustar\ ToO observations (Van Kooten et al. in prep.). We thus investigated the ToO spectral properties to check for any notable radiative variability between the two epochs. We simultaneously fit the ToO phase-averaged spectra with the \nustar+\xmm\ spectra, while allowing all spectral parameters, except for the absorption column, free to vary. The resulting parameters agree within $\sim16\%$ and are consistent within their statistical uncertainties, thus we deem any significant long-term variability due to the glitch unlikely given the large calibration uncertainties due to the ToOs being taken close to the sun. 



\subsubsection{Phase-Averaged Spectroscopy} 
\label{sec:phaseavg_spec}

\begin{figure}[t!]
    \hspace{-0.75cm}
    \includegraphics[width=\linewidth]{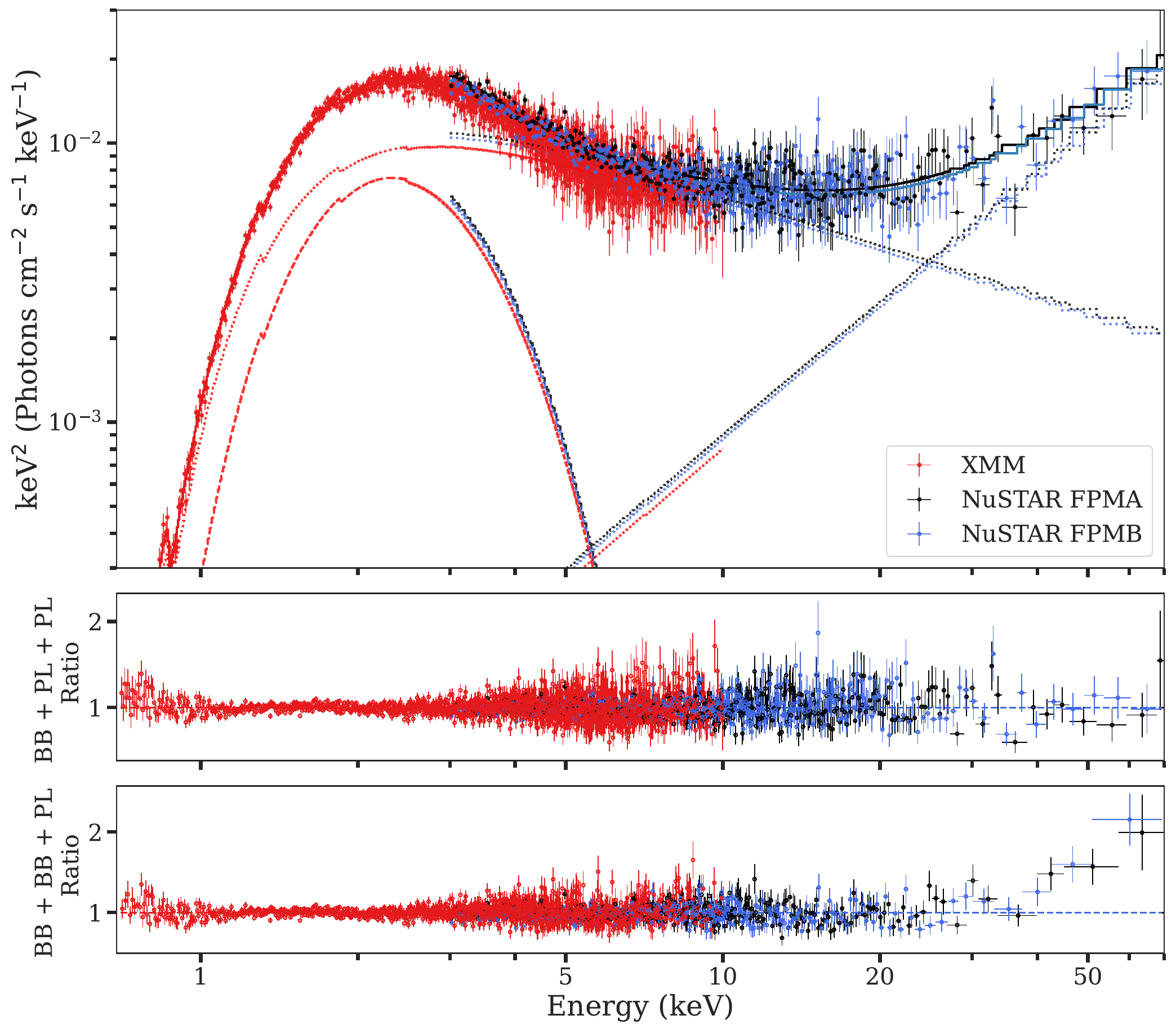}
    \caption{{\sl Top panel.} $\nu F_{\nu}$ phase-averaged spectra of \src\ as observed with \xmm, shown in red, and \nustar\ FPMA and FPMB shown in black and blue, respectively. The solid lines represent the best-fit absorbed BB+2PL model while the dashed lines depict the best-fit of each additive model component. The points show the spectral data with the folded model and their respective uncertainties. {\sl Middle panel.} Ratio of the data to the best-fit BB+2PL model. {\sl Bottom panel.} Ratio of the data to the best fit 2BB+PL model. See text for more details.}
    \label{fig:phase_averaged_spectra}
\end{figure}

We fit the simultaneous \nustar+\xmm\ spectra in the 3--70~keV and 0.5--10~keV range, respectively, with an absorbed blackbody (BB) and double power-law (PL) model. This model results in a satisfactory fit with a C-stat of 3584 for 3413 degrees of freedom (d.o.f). The data and the spectral model in $\nu F_\nu$ space are shown in the upper panel of Figure \ref{fig:phase_averaged_spectra}, while the corresponding residuals, i.e., the ratio of data divided by the folded model, are shown in the middle panel. We also fit the data with an absorbed 2BB and PL model with a C-stat of 3648 for 3413 d.o.f. This is a $\Delta$C-stat of 64 for the same number of model parameters as above. The residuals of this fit are shown in the lower panel of Figure \ref{fig:phase_averaged_spectra}, where large residuals are apparent at energies $>$40~keV. We deduce that a BB+2PL model better represents the simultaneous, phase-averaged \nustar+\xmm\ broadband spectra. We utilized this model for all our subsequent analyses unless we state otherwise.


We find the best fit neutral hydrogen column density, $N_{\rm H}$,  to be $(1.89\pm 0.02) \times 10^{22}$ cm$^{-2}$. We also find a BB temperature, $kT = 0.468 \pm 0.003$ keV and a BB thermal emission area of $R^2 = 8.2 \pm 0.4$ km$^2$ assuming a distance of 3.85 kpc \citep{2006ApJ...650.1070D}. Additionally, we obtain values for the soft PL photon index $\Gamma_{\rm soft}$ of $ 2.63 \pm 0.04$ and the hard PL photon index $\Gamma_{\rm hard}$ as $0.5 \pm 0.1$. 

We used the convolution model \texttt{cflux} within Xspec to estimate the unabsorbed fluxes at different energy ranges as well as for the thermal and nonthermal spectral components. We calculate the 0.5--70 keV total unabsorbed flux to be $F_{\rm tot}=(1.48 \pm 0.01)\times 10^{-10}$ erg s$^{-1}$ cm$^{-2}$.
We also estimate the 0.5--10 keV and 10--70 keV fluxes to be $F_{\rm 0.5-10 keV}=(1.17 \pm 0.008)\times 10^{-10}$ erg s$^{-1}$ cm$^{-2}$ and $F_{\rm 10-70 keV}=(3.04 \pm 0.07)\times 10^{-11}$ erg s$^{-1}$ cm$^{-2}$, respectively.

We compute a BB flux of $F_{\rm BB}=(2.81 \pm 0.04)\times 10^{-11}$ erg s$^{-1}$ cm$^{-2}$, a soft power law flux of $F_{\Gamma_{\mathrm{soft}}} = (9.66 \pm 0.08)\times 10^{-11}$ erg s$^{-1}$ cm$^{-2}$ in the 0.5--70 keV range, and a hard power law flux of $F_{\Gamma_{\mathrm{hard}}} = (1.95 \pm 0.08)\times 10^{-11}$ erg s$^{-1}$ cm$^{-2}$ in the 3--70 keV range.
Table \ref{table:phase_averaged_and_phase_resolved_parameters} contains the values for all the fitted phase-averaged BB+2PL spectral components and fluxes.

\subsubsection{Phase-Resolved Spectroscopy}
\label{sec:prs}


\begin{figure*}[t!]
    \hspace*{+2.25cm}
    \includegraphics[width=0.75\linewidth]{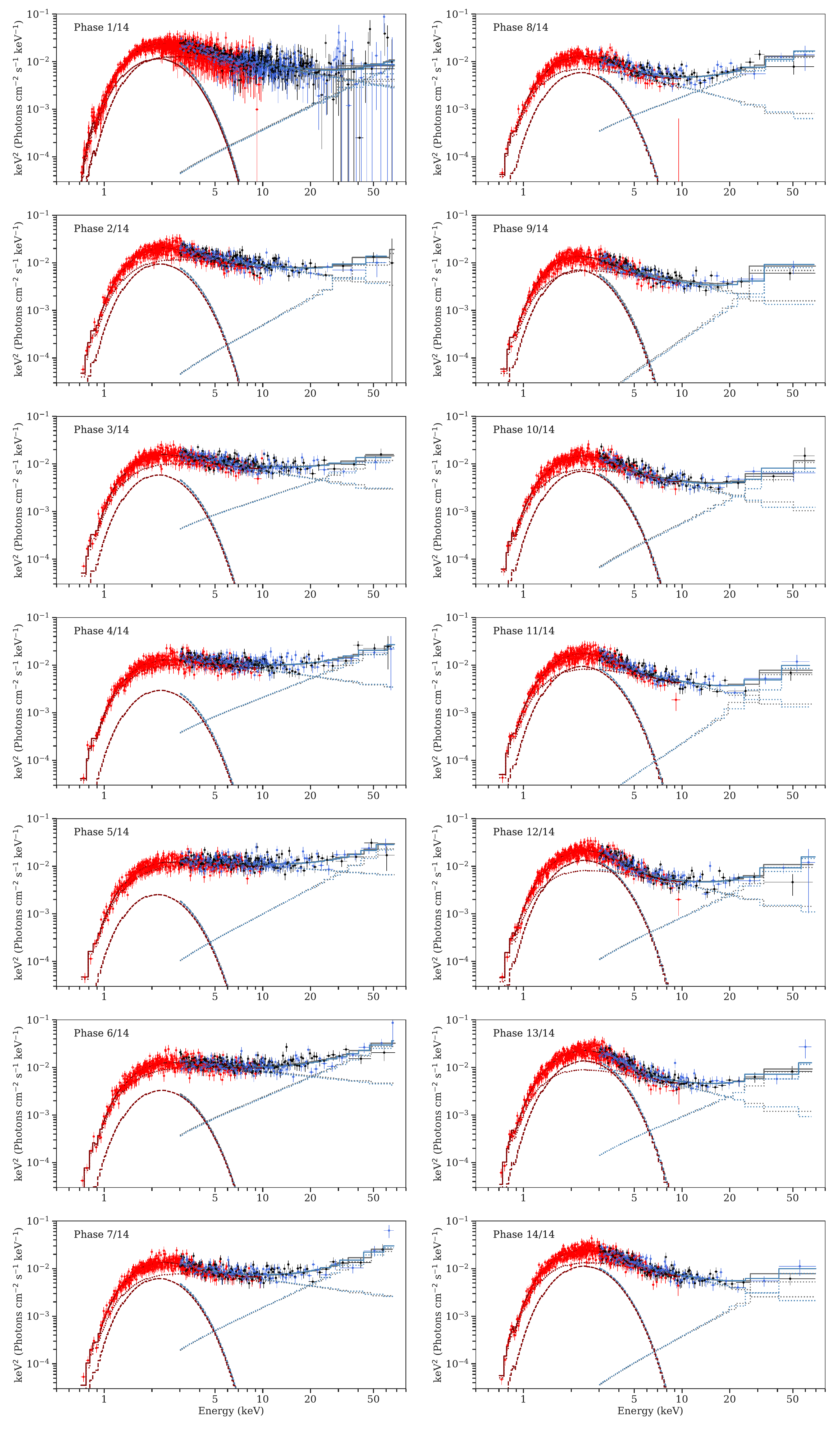}
    \caption{$\nu F_{\nu}$ spectra of 14 phase-resolved bins  of J1708 with \xmm\ shown in red and \nustar\ FPMA \& FPMB shown in black and blue respectively with a BB+2PL model applied. Two \nustar\ ToO observations were used to improve the statistics of the model fitting, however, for visual simplicity they are not shown above. }
    \label{fig:phase_resolved_spectra}
\end{figure*}

\begin{figure*}[t!]
    \includegraphics[width=1.0\linewidth]{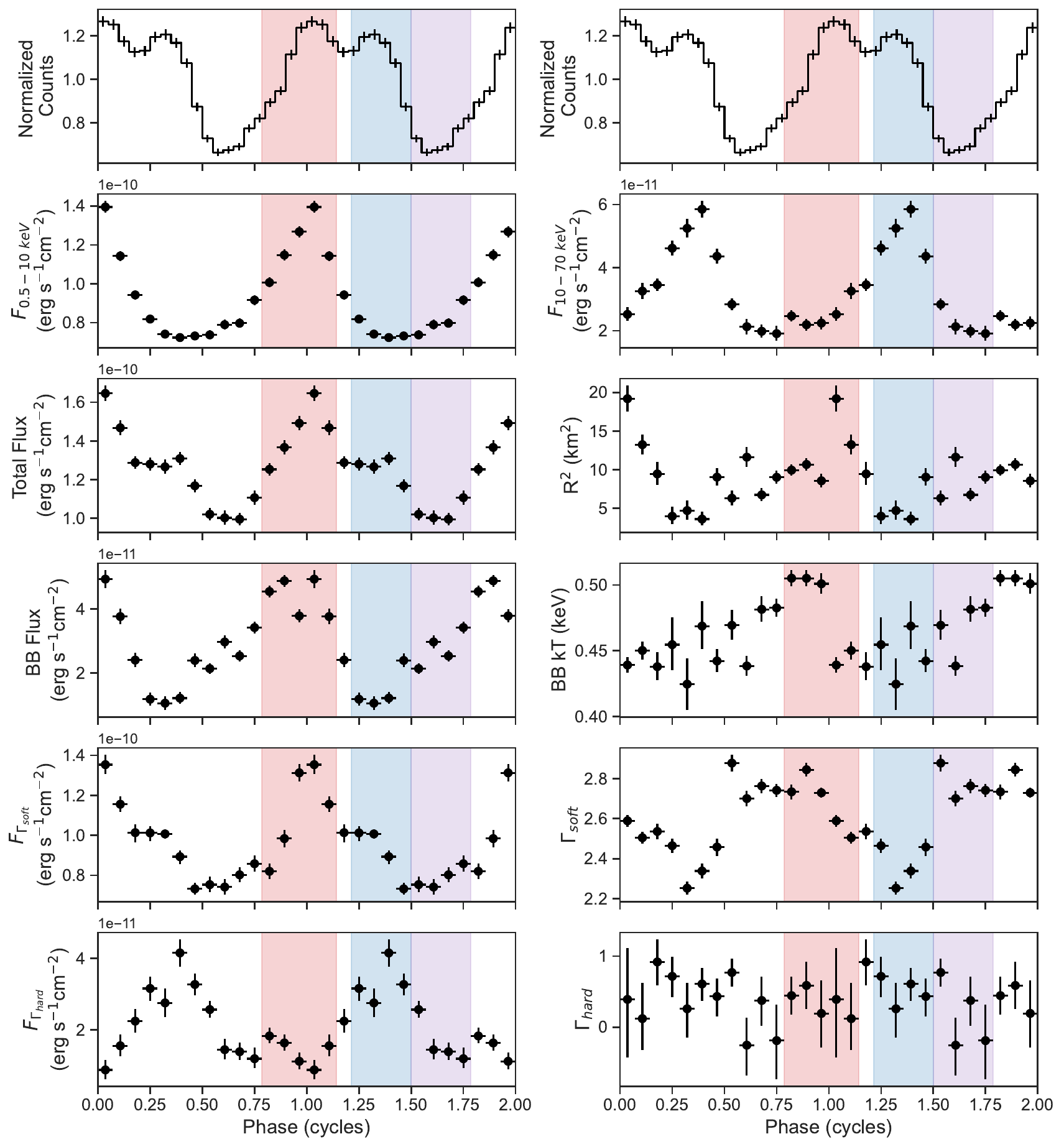}
    \caption{Phase-resolved spectral parameters from an absorbed BB+2PL Xspec model used to fit \nustar+\xmm\ spectral data of \src~ divided into 14 phase-bins. The top panels show the repeated \nustar\ 3--70 keV normalized intensity pulse profiles for reference. The vertical bands correspond to the soft peak (red), hard peak (blue), and the off peak (purple) of the pulse profile. In descending order, the left column displays the pulse profile, total 0.5--10 keV unabsorbed flux, total 0.5--70 keV unabsorbed flux, the flux associated with the BB component of our spectral model, the flux of the soft power law component, and lastly the flux of the hard power law component. Likewise (in descending order), the right column displays the pulse profile, total 10--70 keV unabsorbed flux, blackbody emission $R^2$, BB temperature, the soft power law index, and the hard power law index. All fluxes are in units of  erg s$^{-1}$ cm$^{-2}$. See text for more detail. 
    }
    \label{fig:spectralparams}
\end{figure*}

\begin{figure*}[ht!]
    \includegraphics[width=1.0\linewidth]{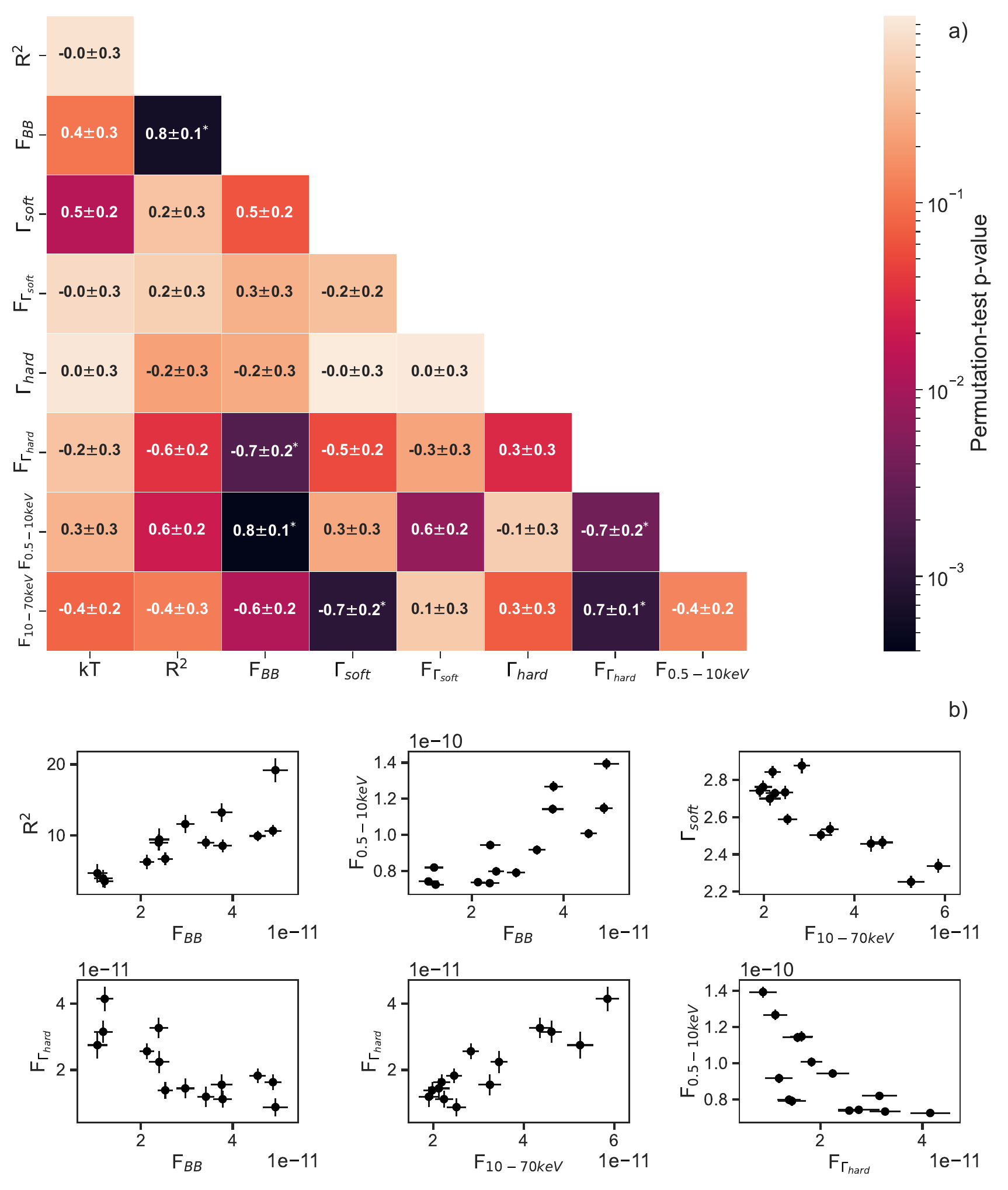}
    \caption{Panel a: Spearman-Rank correlation matrix of the phase-resolved BB+2PL spectral parameters. The colorbar
displays the log-scale of p-values. A single asterisk indicates values with significance levels exceeding 3$\sigma$. Panel b: Plots of correlated parameters with significance higher than 3$\sigma$ as shown in the correlation matrix.}
    \label{fig:correlationmatrix}
\end{figure*}

We performed phase-resolved spectroscopic analysis of \src\ utilizing the simultaneous \nustar+\xmm\ observation along with the two phase-aligned \nustar\ ToOs; we divided the resultant pulse profile into 14 equal phase bins. This binning scheme was selected in order to increase the S/N per spectra, especially at high energies, while adequately sampling the two main pulse peaks in the profile. We fit the spectrum of each phase bin with an absorbed BB+2PL model. We fix the hydrogen column density $N_{\rm H}$ to the phase-averaged best fit value of $1.89 \times 10^{22}$ cm$^{-2}$. This model resulted in a satisfactory fit for all the spectra. We present the phase-resolved $\nu F_\nu$ spectral data along with the corresponding best fit model in Figure \ref{fig:phase_resolved_spectra}, and the spectral parameters and corresponding fluxes are shown in Table \ref{table:phase_averaged_and_phase_resolved_parameters}. 

It is evident that \src\ exhibits strong variability with phase, in spectral shape and normalization. For instance, the soft PL becomes noticeably more flat in phase bins 4-7, while on the other hand, in phase bins 11 to 13, the soft X-ray PL is comparatively steep while the BB component dominates the soft X-ray emission.



For a more comprehensive visualization of the phase-dependence of the spectral variation, we display in Figure~\ref{fig:spectralparams} the separate spectral parameters as a function of phase. Namely, we show the 0.5--10~keV, 10--70~keV, and 0.5--70~keV fluxes; the BB temperature $kT$, area $R^2$ (assuming a source distance of 3.85 kpc; \citealt{2006ApJ...650.1070D}), and its flux; and the photon indices and fluxes of the soft and hard PL components, respectively. Finally, the top row shows the 3--70 keV normalized-count pulse profile (PP). Each profile displays two cycles for clarity.

We find a well-defined pulse in the soft X-ray flux, $F_{0.5-10\mathrm{ keV}}$, that distinctly aligns with the first peak of the energy-integrated PP. This peak is dominated by the emission from the BB and soft PL components. We refer to this peak as the `soft peak' (denoted by the red band in Figure \ref{fig:spectralparams}). The 10--70~keV flux, dominated by emission from the soft and hard PL components, also exhibits a clear single peak, which we refer to as the `hard peak' (blue band in Figure \ref{fig:spectralparams}). Spectrally, the BB component shows a hint of $\approx15\%$ increase in temperature, yet an emitting area a factor $\sim5$ larger at the soft X-ray peak (i.e., in the red band) compared to its off-pulse (in the blue band). The soft X-ray PL exhibits prominent curvature variability, showing that the hardest spectra align with the hard peak. On the other hand, the hard X-ray PL does not display strong phase-variability, though this might be due to the associated large uncertainties which are on average at the $50\%$ level. Finally, we relied on the soft and hard X-ray flux pulse profiles to define an off-pulse region, shown as the purple band in Figure \ref{fig:spectralparams}. We note, however, that the BB flux is rising during these phase bins\footnote{Given the complexity of the profile, it is non-trivial to define an off-pulse region where all pulse-peaks are at a minimum.}.

We estimate the uncertainty and significance of the correlation between the spectral parameters presented in Figure \ref{fig:correlationmatrix} (panel a) using Monte Carlo bootstrap composite technique (see e.g. \citealt{laken_sw_2013}). For each set of assessed parameter pairs, we resampled the data with replacement 10,000 times to produce the bootstrap and randomly perturb each resampled realization using its measurement uncertainty (we assume Gaussian errors). Then, we calculated the Spearman-Rank correlation coefficient for each bootstrapped data set where the mean is the reported coefficient value and the uncertainties are given by the standard deviation of the bootstrapped set. We evaluate the statistical significance of the correlations by calculating the two-sided p-values (under a null hypothesis of no correlation) using a permutation test with 10,000 random permutations. The value of the colorbar displays the log-scale of the p-values associated with the correlations, where the correlations with higher significance levels are darker in color. Correlations with a statistical significance above the 3$\sigma$ level are noted with a single asterisk.


The correlation plots with statistical significance at or exceeding the 3$\sigma$ level are presented in Figure \ref{fig:correlationmatrix} (panel b). Most notably is the strong anti-correlation between the $F_{10-70 \mathrm{keV}}$ flux and the soft PL photon index $\Gamma_{\mathrm{soft}}$. We also find a strong correlation between the BB flux, $F_{\rm BB}$, and $R^2$. The rest of the correlations are those related to the fluxes of the individual components, as well as the soft $F_{0.5-10 \mathrm{keV}}$ and hard $F_{10-70 \mathrm{keV}}$ X-ray fluxes. Namely, the BB flux, $F_{\mathrm{BB}}$, correlates with $F_{0.5-10 \mathrm{keV}}$, and anti-correlates with $F_{\Gamma_{\mathrm{hard}}}$, whereas $F_{0.5-10 \mathrm{keV}}$ correlates with $F_{\Gamma_{\mathrm{soft}}}$ and anti-correlates with $F_{\Gamma_{\mathrm{hard}}}.$ Lastly, $F_{10-70 \mathrm{keV}}$ displays a strong correlation with $F_{\Gamma_{\mathrm{hard}}}$.

\begin{figure*}[t!]
    \includegraphics[width=\linewidth]{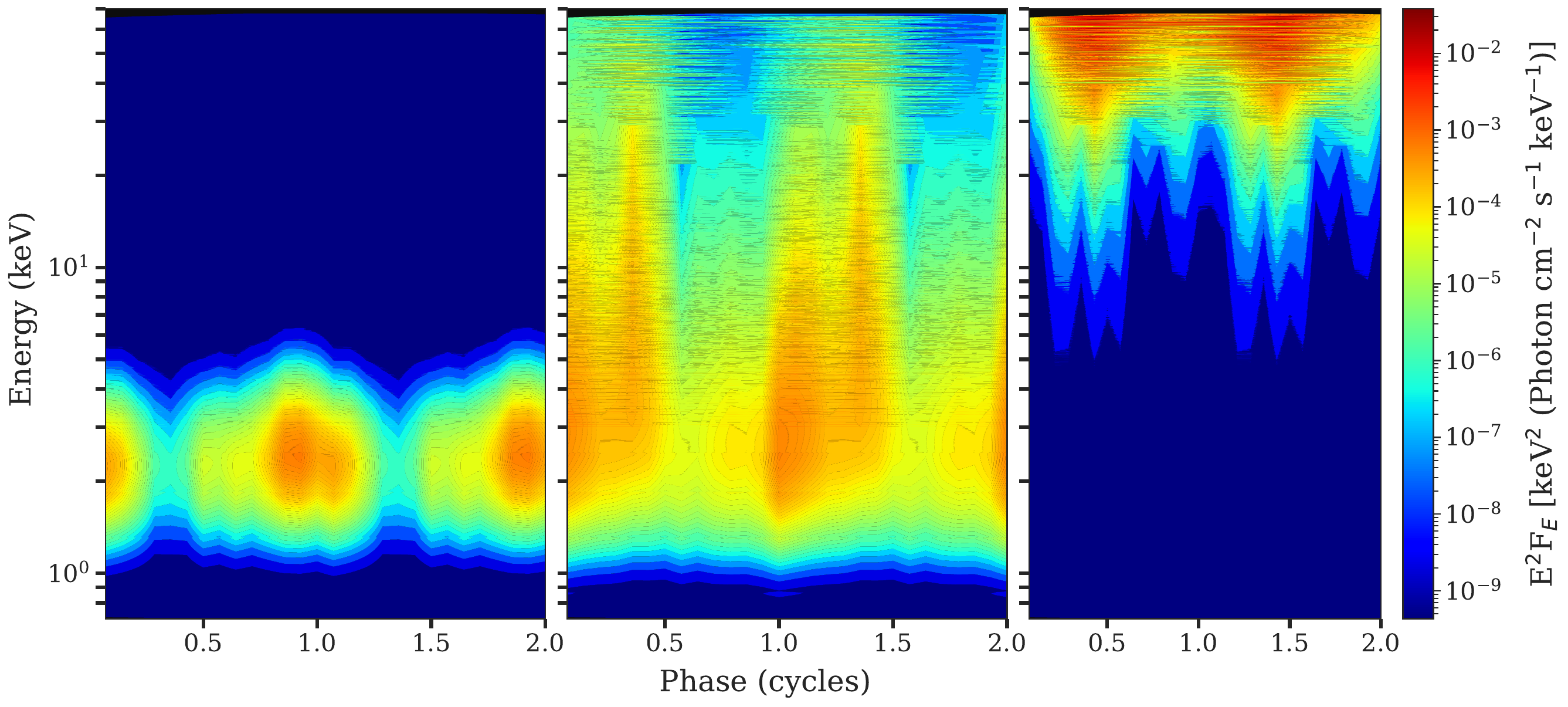}
    \caption{Phase-energy diagram of BB+2PL spectral model of \src\ between 0.5--70 keV. The left panel shows the BB $E^2F_E$ component, the soft PL $E^2F_E$ is displayed in the center, and the hard PL $E^2F_E$ component is on the right. The colorbar displays the power of the $E^2F_E$.}
    \label{fig:dynamicspectralprofiles}
\end{figure*}

Another way to examine the phase-dependent evolution of the broadband spectra is with dynamic spectral profiles (DSPs), shown in Figure~\ref{fig:dynamicspectralprofiles}. The DSP displays the contour plot of the photon flux $E^2F_E$ (in units of keV cm$^{-2}$ s$^{-1}$) for each spectral component as a function of phase and energy according to the same 14 phase bins described above. The BB $E^2F_E$ plot is presented in the left panel, the soft PL in the middle panel, and the hard PL in the right panel of Figure~\ref{fig:dynamicspectralprofiles}, respectively. The intensity scales according to the colorbar to the right of the figure.

The BB flux is most prominent at soft X-ray energies up to approximately 4 keV and exhibits a single broad-peak centered around 0.9 cycles, with an apparent leading shoulder. The $F_{\Gamma_{\mathrm{soft}}}$ displays a double-peaked profile. Its initial peak near $\phi=1.0$ cycles overlaps with the BB contribution until it dominates at $\gtrsim4$~keV. This soft PL peak is slightly offset from the BB maximum and shows a subtle energy-dependent phase shift; the peak drifts rightward with increasing energy. The secondary peak at $\phi \approx 1.35$ cycles is narrow in phase and extends towards higher energies without any apparent energy-dependent phase shifts. This component intersects the $F_{\Gamma_{\mathrm{hard}}}$ at $\approx$ 10--20 keV. At higher energies (20--30~keV), the $\Gamma_{\mathrm{hard}}$ flux becomes dominant with a sharp increase in power (due to the inverted nature of the spectrum). The $F_{\Gamma_{\mathrm{hard}}}$ maximum occurs at roughly $\phi\sim0.4$ cycles, close to the minima of $F_{\mathrm{BB}}$ and the secondary-peak maximum of the soft PL.


\subsection{Polarimetric and Spectro-Polarimetric Analysis}


We simultaneously fit the phase-averaged \nustar+\xmm\ spectra with the \ixpe\ Stokes I, Q, and U spectra in Xspec using a \texttt{tbabs*(bbodyrad*polconst+pow*polconst+pow)} model. We note that since the hard PL component emerges outside of the IXPE energy range, no polarization data was available for this component and thus
no polarization component assigned to it. We link the model parameters between all spectra given the lack of any significant short or long-term variability in the source from RXTE and NICER monitoring through 2025 \citep[e.g., Section~\ref{sec:phaseavg_spec} and][Van Kooten et al. in prep]{2014ApJ...783...99S}. We group the \nustar+\xmm\ spectra to have a minimum of 5 counts per bin and fit the spectral models using the W-statistic while for the \ixpe\ observation we grouped according to at least 25 counts per spectral bin and employed the Xspec \texttt{statistic chi}. The fit is statistically adequate with a total fit statistic of 4988.75 for 4759 d.o.f. Table \ref{table:spectral_decomposition} summarizes the spectro-polarimetric best fit model parameters. 

We find a spectrally-decomposed polarization degree (PD) and angle (PA) of $\text{PD}_{\text{BB}} = 0.51 \pm 0.08$ and $\text{PA}_{\text{BB}} = 31 ^{\circ} \pm 4$ for the BB component, and $\text{PD}_{\Gamma_{\text{soft}}} = 0.84 \pm 0.05$ and $\text{PA}_{\Gamma_{\text{soft}}} = -61^{\circ} \pm 1$ for the soft PL. 
Note the $\sim 90^{\degree}$ swing between the $\text{PA}_{\text{BB}}$ and $\text{PA}_{\Gamma_{\text{soft}}}$. Our fit parameters are largely in agreement with those presented in \citet{2023ApJ...944L..27Z}, yet with a somewhat improved constraint on the PD of the BB component.

Next, we examine the phase-averaged polarization properties of \src\ as a function of energy using \texttt{ixpeobssim} \citep{2022SoftX..1901194B}. We bin the data into five energy bands, namely 2--3, 3--4, 4--5, 5--6, and 6--8 keV. We find a roughly constant PA across energies at $\approx -60^{\degree}$ while the PD increases as a function of energy from $\sim 20\%$ at 2--3 keV up to $\sim 60 \%$ at 6--8 keV.
These results are fully consistent with the analysis presented in \citet{2023ApJ...944L..27Z}.



\begin{figure*}[ht!]\label{phase_resolved_PA_PD}
    \hspace{-0.75cm}
    \includegraphics[width=1.1\linewidth]{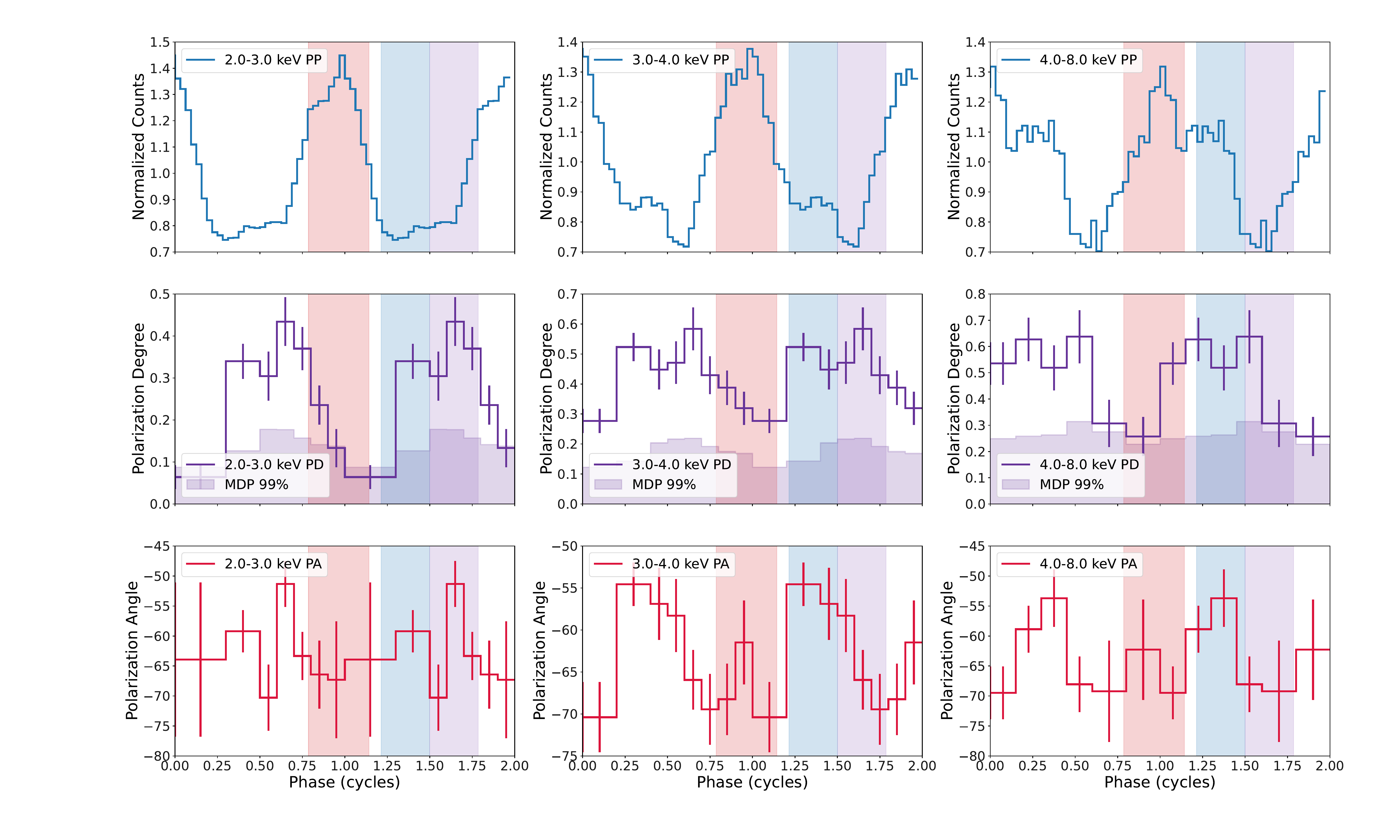}
    \caption{\ixpe\ energy-resolved pulse profiles (top row), polarization degree (middle row) and polarization angle (bottom row) of \src\ binned inhomogeneously according to three energy bands, i.e., 2--3 (left), 3--4 (center), and 4--8 keV (right). The soft peak of the energy-integrated (0.5--70 keV) pulse profile is denoted by the light red vertical band, the location of the hard peak is marked by the light blue band, and the off peak is shown in purple (see also Figure \ref{fig:spectralparams} top row). The corresponding MDP$_{99\%}$ for each given energy range is overlaid atop the PD profiles.} 
    \label{fig:PA_PD_phases}
\end{figure*}

Finally, we performed detailed energy- and phase-dependent polarization analyses. We derive the PA and PD according to the inhomogeneous phase bins that optimize the resolution while remaining statistically significant (i.e., exceed the minimum detectable polarization at the 99\% confidence level, MDP$_{99\%}$). We then examined the PA and PD phase diagrams in the 2--3, 3--4, and 4--8 keV energy ranges. These polarization results along with the corresponding \ixpe\ intensity pulse profiles for the specified energy ranges are presented in Figure \ref{fig:PA_PD_phases}. 


The PD and PA exhibit variation in phase across the energy bands. In the 2--3 keV band and 3--4 keV band, the PD is anti-correlated with the main peak of the intensity profile. The PD of the 2--3 keV band reaches a minimum consistent with the MDP$_{99\%}$ and a maximum of $43\pm6\%$, while for the 3--4 keV band, the PD reaches a maximum of $58\pm6\%$ and a minimum of $28\pm4\%$. In both energy bands, the PD is significantly detected in the phases consistent with emission from the soft PL non-thermal component (blue band). The 4--8~keV band exhibits a maximum PD=$64\pm10\%$ that is phase-aligned with the non-thermal PL peak, and a minimum of $28\pm4\%$ coincident with the leading edge of the thermal component (first bin in the red band). We note that the trailing edge of the thermal component peak exhibits the lowest PD in the 2--3~keV band, while increasing in the 3--4~keV band, and becoming part to the largest polarized emission with rotational phase in the 4--8~keV band, in tandem with the emergence of the non-thermal component at those phases. The PA in the 2--3 keV band possesses large errors bars that make discerning a demonstrable phase-dependence difficult. In the 3--4 and 4--8 keV bands, the PA displays clear phase-variation between $\approx-70^{\degree}$ and $-55^{\degree}$. We find that the phase-dependent polarimetric characteristics trends are largely consistent with the phase- and energy-resolved behavior reported by \cite{2023ApJ...944L..27Z}. use of a different phase and energy binning scheme, which was chosen to facilitate comparison with the contemporaneous NuSTAR and XMM-Newton spectral properties.


Lastly, we attempted to perform phase-resolved spectro-polarimetry to derive the PA and PD of the separate spectral components, yet the fits were not well-constrained due to the low statistics.

\subsection{Physical Modeling}\label{sec:modeling}
\subsubsection{Atmospheric Modeling}

\begin{figure*}[t!]
    \centering
\includegraphics[width=1\linewidth]{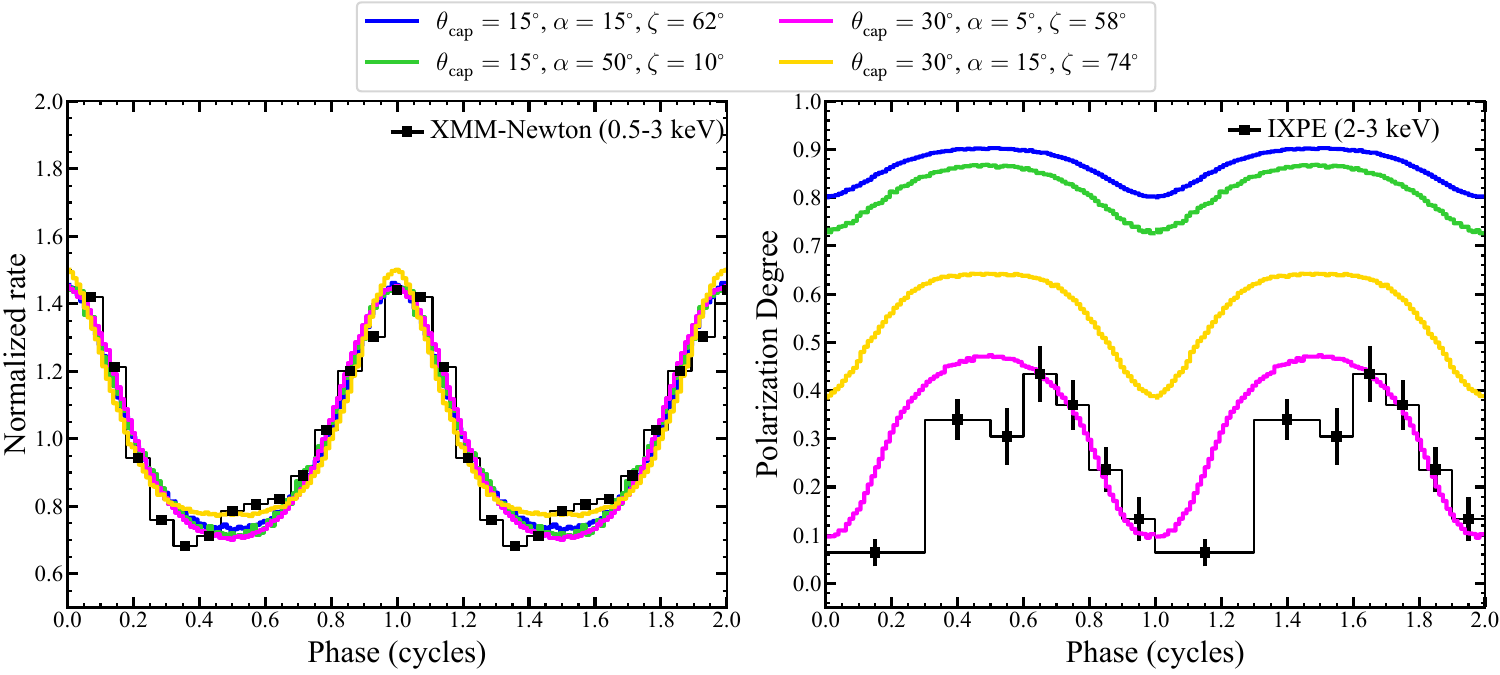}
    \caption{Simulated intensity (left) and PD (right) pulse profiles from {\sl MAGTHOMSCATT} for two antipodal polar caps extending from the respective magnetic poles to colatitudes of  $\theta_{\rm cap} = 15^{\circ}, 30^{\circ}$ as a function of rotational phase in units of cycles $\Phi / (2\pi)$. The black histogram on the left (right) panel represents the intensity (PD) data extracted from \xmm\ (\ixpe) in the energy of 0.5--3 keV (2--3 keV).  Four solid curves on the left panel represent four different parameter sets of ($\theta_{\rm cap}, \alpha, \zeta$) that give a good match to the observed count rates. The traces on the right panel show the corresponding polarization pulse profiles, wherein magnetospheric birefringent propagation of light is neglected.}
    \label{fig:MTS-fit-J1708}
\end{figure*}

Pulse profile shapes in both intensity and now polarization provide important constraints on magnetar geometry and surface emission locales.  For intensities, this has been a practice over the last two decades, being employed by several groups to probe the size of the hot emission region, its locale on the surface, the value of the star's magnetic inclination angle \teq{\alpha} in a dipolar field morphology, and the observer's viewing direction angle \teq{\zeta}, both relative to rotation axis \teq{\boldsymbol{\Omega}}; see, e.g., \cite{Albano-2010-ApJ,Bernardini-2011-MNRAS,  2020ApJ...889L..27Y}.   Adding information on polarized X-rays detected by \ixpe\ from magnetars enables a new dimension in diagnostic space.

To interpret our analysis of the intensity and \ixpe\  polarization data for \src, we present here some atmospheric model results from the magnetic Thomson Monte Carlo simulation {\sl MAGTHOMSCATT}, whose construction and application are detailed in \cite{Barchas-2021-MNRAS} and \cite{Hu-2022-ApJ, Dinh-2025-ApJ, Dinh-2026-ApJ}.  
Extended locales for the emitting surface were adopted.  These were uniformly hot, antipodal polar caps extending from each pole to some colatitude \teq{\theta_{\rm cap}} away. Slab locales for each photon are randomly selected throughout these caps. After the atmospheric transport, the radiation was then propagated through a Schwarzschild metric, specified by a neutron star of mass \teq{\mns = 1.44 M_{\odot}} and radius \teq{\rns = 10^6}cm, to observers sampling all points on the sky.

In our simulation runs, there are three main geometrical parameters, namely, $\alpha,\ \zeta,\ \theta_{\rm cap}$.
For each of two choices \teq{\theta_{\rm cap}=15^{\circ}, 30^{\circ}}, we performed a suite of runs for different \teq{\alpha} values, and then compared the simulated pulse profiles with the corresponding observed data below $3$~keV.
We focused on this restricted energy range because the spectroscopy performed above strongly suggests that this is the range that predominantly captures surface thermal signals. Emergent directions on the sky at infinity correspond to different rotational phases \teq{\Phi = \Omega t}, so that the 2D ``sky maps'' in \teq{\Omega t - \zeta} space are generated by the simulation runs, as detailed in \cite{Hu-2022-ApJ}. Constant \teq{\zeta} cuts of these maps deliver pulse profiles of intensity and polarization Stokes parameters for comparison with data.

Most light curves in our \teq{\alpha ,\zeta} sampling do not well describe the intensity pulse profiles exhibited in Figures \ref{fig:nuxmm_energy_dependent_pp} and \ref{fig:spectralparams}. In Figure~\ref{fig:MTS-fit-J1708}, left, we depict four model profiles that do closely match the intensity data overall. We note that the results presented here are solely illustrative, rather than an accurate fit to the data. The parameter variation in these preferred models is quite broad, spanning nearly aligned (\teq{\alpha = 5^{\circ}}) and moderately oblique rotators (\teq{\alpha = 50^{\circ}}), and indicates that the magnetar stellar geometry and emission region size cannot be tightly constrained by intensity information alone, a conclusion obtained in the study of \cite{Dinh-2026-ApJ}.

Polarization data can help break such degeneracies, and so we generated linear PD light curves for the same four models. These are displayed in Figure~\ref{fig:MTS-fit-J1708}, right, with polarization propagation through the magnetosphere being restricted to purely general relativistic parallel transport \citep{Hu-2022-ApJ}.  This added constraint suggests that the \teq{\theta_{\rm cap}=30^{\circ}}, \teq{\alpha = 5^{\circ}} is preferred over the other three, yet it is still not an excellent match to the observed PD data.

This brief presentation omits two key polarization elements germane to magnetar studies.  The first is how the magnetosphere acts as a polarimetric crystal due to the role of QED magnetic vacuum birefringence on the transport of polarizations to infinity.  This effect preserves the polarization eigenstates during photon propagation out to high magnetospheric altitudes, that is, preserves the high PD emergent at the surface, hence enhancing the net observed PD \citep{Heyl-2000-MNRAS,Heyl-2002-PRD}. This influence is detailed in \cite{Dinh-2026-ApJ}, wherein phase-dependent linear PD above 0.8 are routinely realized for 1RXS~J1708-40 using the similar atmosphere simulation model output as displayed in Figure~\ref{fig:MTS-fit-J1708}.  The second element is that of polarization mode switching deep within the atmosphere \citep{Lai-Ho-2003-ApJ,Lai-Ho-2003PhRvL,  Lai-2023-PNAS, 2024MNRAS.528.3927K}, which arises due to the competition between QED vacuum and plasma dispersion at higher atmospheric densities.  This generally depolarizes the signal from the surface relative to the high local values that are generated from {\sl MAGTHOMSCATT}, thereby lowering the net enhancement due to the magnetospheric QED birefringence.   Combining the competing influences of atmospheric mode switching and magnetospheric birefringent propagation is an involved task deferred to future work, and is likely to generate a broad range of net PD when varying the main model parameters \teq{\alpha,\ \zeta}, and \teq{\theta_{\rm cap}}.

\subsubsection{Non-thermal Modeling}

\begin{figure} 
\centerline{\includegraphics[width=10cm]{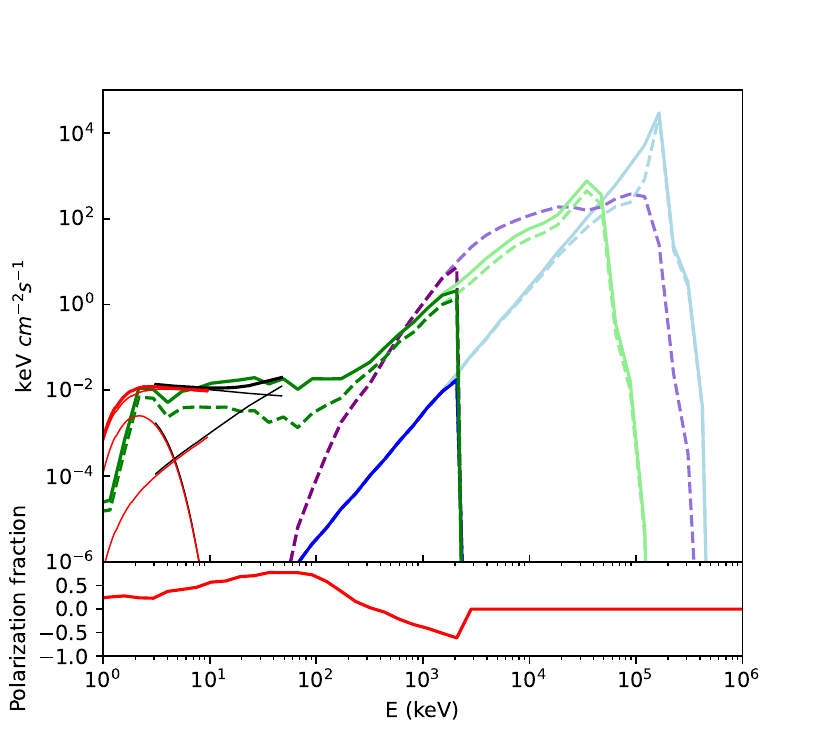}}
\caption{Model spectra and polarization from a pair cascade initiated by electrons with constant Lorentz factor $\gamma = 10^3$ on a closed dipole field loop with maximum radius, $r_{\rm max} = 6$ neutron star radii, for observer angle to the magnetic pole, $\theta_v = 108^\circ$ and a surface magnetic field, $B_0 = 4.4 \times 10^{14}$ G.  Top panel: Photon spectral energy distributions of primary resonant inverse Compton scattering (blue), photon splitting (purple) and pair synchrotron radiation (green) for $\parallel$ (dashed lines) and $\perp$ (solid lines) photon polarization.  Light lines/dark lines are spectra unattenuated/attenuated by pair production and photon splitting.  Also shown are the \xmm\ (red) and \nustar\ (black) spectral fits for phase interval 5 (see Figure \ref{fig:phase_resolved_spectra}). Bottom panel: Polarization fraction, $(\perp - \parallel)/(\perp + \parallel)$, as a function of photon energy for the total spectra above.}
\label{fig:cascade_spectra}
\end{figure}

The very high degree of polarization observed by \ixpe\ for the soft PL component in the 4--8~keV energy range ($\approx60\%$ at certain rotational phases, Figure~\ref{fig:PA_PD_phases}) could be explained by synchrotron radiation.  Such a synchrotron component can be generated by radiation from electron-positron pairs that are produced in cascades on closed magnetic field loops.  Particles accelerated on  twisted field loops resonantly scatter \citep{wadiasingh18ApJ} thermal photons from the hot neutron star surface to form a broad spectrum of photons that undergo pair production and photon splitting in the strong magnetic field \citep{Harding2025ApJ}. In the strongest fields near the neutron star surface, the pairs are produced in very low Landau states, radiating only a few high energy photons.  At higher altitudes pairs are produced in higher Landau states, generating a broader spectrum of lower-energy photons.  At certain viewing angles, both the high energy and low energy synchrotron components are visible.  

Figure \ref{fig:cascade_spectra} shows the spectrum of a cascade simulation where both pair synchrotron components (in green) are present.  The synchrotron spectrum breaks around an energy of 30--50 keV, with the hard spectrum above the break produced by first-generation pairs in low Landau states and the softer spectrum below the break produced by higher-generation pairs in high Landau states.  The PD of the soft component reaches near 60\% by 8 keV, dominated by X-mode ($\perp$) photons, and could explain the observed soft power law spectral and polarimetric properties. The harder component reaches an even higher degree of polarization and could account for the observed hard power law emerging above 30 keV. The split photon spectrum begins to dominate above $\sim 600$ keV and reverses the sign of polarization as splitting in the $\perp \rightarrow \parallel\parallel$ mode produces only O-mode ($\parallel$) photons.  All spectra are attenuated around 3 MeV by pair production (for details of the cascade calculation, see \citealt{Harding2025ApJ}).  The \xmm\ and \nustar\ BB and PL fits for phase interval 5 in Figure \ref{fig:phase_resolved_spectra}, where the non-thermal components dominate, are also shown in Figure \ref{fig:cascade_spectra}.  

The phase-resolved analysis shown in Figures \ref{fig:phase_resolved_spectra} and \ref{fig:spectralparams} shows that the soft power law photon index varies with phase, with a harder spectrum near the flux peaks of the non-thermal components and a softer spectrum toward the BB flux peak and in the off-peak phase.  The model phase-resolved spectra also show a maximum hardness of the soft synchrotron component at the pulse peak and a softer spectrum in the pulse wings.  However, the index of the hard synchrotron component does not vary with phase.  The spectral index of the hard power law in Figure \ref{fig:spectralparams} similarly does not show a significant phase variation although the error bars are larger.

\section{Summary \& Comparison with Previous Works}{\label{section4}}

We have reported the results of highly detailed temporal, spectral, and spectro-polarimetric analysis performed upon the magnetar 1RXSJ170849.0--400910 using \nustar, \xmm, and \ixpe\ data. We present the characteristics of the broadband 0.5-70 keV, energy-resolved X-ray pulse morphology, phase-averaged and phase-resolved spectral properties, and energy and phase-dependent polarimetric results.

 Figure \ref{fig:energy_integrated_pp} shows the energy and phase-integrated \xmm\ 0.5--10 keV and \nustar\ 3--70 keV pulse profiles. The soft X-ray pulse profile has one dominant peak and a subpeak that becomes more prominent in the 3--70 keV, showcasing a dependence upon energy. The energy-dependent pulse morphology evolution seen in Figure \ref{fig:nuxmm_energy_dependent_pp} is consistent with the \xmm, \textit{INTEGRAL}, and \textit{RXTE-PCA} energy-dependent pulse profiles presented by \citealt{2008A&A...489..263D}. Notably, since the pulse profiles we present possess a finer energy resolution than those reported in \citealt{2008A&A...489..263D}, we can see certain features, such as the shoulder emerging in the soft X-ray (e.g. at $\phi = 0.6$ for 0.5--1.25 keV), in higher detail. Figure \ref{fig:rms_pf_energy_resolved} presents the RMS PF as a function of energy in which we see a fairly constant PF around $\approx 25\%$ until a dip emerges around 3 keV to about $\approx 20\%$. The PF steadily climbs at higher energies until a plateau at $\approx 41\%$ begins forming around 20 keV.  These values are similar to those in \citealt{2008A&A...489..263D} which found an average 2--10 keV PF
of $(35.6 \pm 1.4)$\% with \xmm\ and $(39 \pm 6)$\% for the 20--150 keV band using \textit{INTEGRAL-RXTE}, though our \xmm\ PF is lower. Our energy-resolved PF characteristics are also consistent with their findings, with the PF varying at soft X-rays while largely remaining constant at higher energies. 

We fit the simultaneous \nustar+\xmm\ observation along with two later \nustar\ ToOs with an absorbed BB+2PL spectral model, shown in Figure \ref{fig:phase_averaged_spectra}. We found the best fit values for these model parameters to be a column density of $N_{\rm H} = (1.89 \pm 0.02) \times 10^{22}$ cm$^{-2}$, a BB temperature of $kT = 0.468 \pm 0.003$ keV, a BB
thermal emission apparent area of $R^2 = 8.2 \pm 0.4$ km$^2$ for distance 3.85 kpc \citep{2006ApJ...650.1070D}, soft PL photon
index of $\Gamma_s = 2.63 \pm 0.04$ and hard PL photon index
$\Gamma_h = 0.5 \pm 0.1.$ We also calculate the unabsorbed 0.5--70 keV total flux as $F_{\rm tot}=(1.48 \pm 0.01)\times 10^{-10}$ erg s$^{-1}$ cm$^{-2}$ (see also Table \ref{table:phase_averaged_and_phase_resolved_parameters}). These values are consistent with previous spectral fits of \src\ (see e.g. \citealt{2001ApJ...557...18P},\citealt{2008A&A...489..263D}, \citealt{vigano2013MNRAS}), which supports the established long-term radiative and spectral stability observed in this source \citep{2014ApJ...784...37D, 2014ApJ...783...99S}. 

We also performed broadband, phase-resolved spectroscopy using 14 phase bins (spectra shown in Figure \ref{fig:phase_resolved_spectra}). This is a much finer phase bin resolution than produced in \cite{2008A&A...489..263D}, allowing for detailed phase-diagrams of key spectral parameters (Figures \ref{fig:spectralparams} and \ref{fig:dynamicspectralprofiles}) and in-depth exploration of their correlations (Figures \ref{fig:correlationmatrix}). This work extends the spectral coverage of phase-resolved spectral parameter mapping presented in \cite{2001ApJ...560L..65I}, which focused on the \xmm\ band. In particular, the spectral parameter $F_{\mathrm{BB}}$ exhibit strong correlations with $R^2$ and $F_{0.5-10 \mathrm{keV}}$ and anti-correlation with $F_{\Gamma_{\mathrm{hard}}}$. $F_{0.5-10 \mathrm{keV}}$ exhibits a strong correlation with  $F_{\Gamma_{\mathrm{soft}}}$ and anti-correlation with $F_{\Gamma_{\mathrm{hard}}}$. Lastly, $F_{10-70 \mathrm{keV}}$ shows a strong correlation with $F_{\Gamma_{\mathrm{hard}}}$ and a very strong anti-correlation with $\Gamma_{\mathrm{soft}}.$ 

The large S/N and broad energy coverage of the data also enabled us to create detailed DSP that showcase the intensity distribution as a function of both phase and energy, shown in Figure \ref{fig:dynamicspectralprofiles}. The DSP affirms the complex interplay between the disparate components in the broadband emergent radiation of \src. The BB flux is most relevant at energies $<4$ keV, yet with contribution from the soft PL component which subsequently dominates the flux contribution until about 20 keV. From there the hard PL flux dominates in intensity, possibly suggesting most of the magnetar's flux contribution does indeed come from the hard X-rays as \cite{2008A&A...489..263D} proposes.   


We explored the spectropolarimetric characteristics of the source using data from \ixpe. We used Xspec to simultaneously fit the \textit{XMM, NuSTAR} and \ixpe\ data with a BB+2PL with constant polarization for the BB and the soft PL components. We found the phase and energy-integrated spectrally decomposed PD and PA to be $\text{PD}_{\text{BB}} = 0.51 \pm 0.08$, $\text{PA}_{\text{BB}} = 31 ^{\circ} \pm 4$, $\text{PD}_{\text{PL}_{\text{soft}}} = 0.84 \pm 0.05$, and $\text{PA}_{\text{PL}_{\text{soft}}} = -61^{\circ} \pm 1$. We find these phase-averaged polarimetric values to be consistent with those presented in \cite{2023ApJ...944L..27Z} except for the value of $\text{PD}_{\text{PL}_{\text{soft}}}$ which was better constrained in our fit.


Lastly, we investigated the variation of the polarimetric properties as a function of both energy and phase in Figure \ref{fig:PA_PD_phases}. We performed detailed phase-dependent polarization analysis according to an inhomogeneous binning scheme that probes the structure of the PD and PA profiles and their connection to the phase-resolved spectral components in the 2--3, 3--4, and 4--8 keV band. We find the trending PD increases as a function of energy, which aligns with the phase-averaged and phase-dependent behavior reported in \cite{2023ApJ...944L..27Z}, though we adopt a phase and energy binning scheme motivated by the goal of investigating the polarimetric characteristics in the context of the NuSTAR+XMM spectral properties. The PD exhibits strong variation in phase across all energies, while the PA also varies prominently in the 3--4 and 4--8 keV bands.

\section{Discussion}\label{sec:discussion}
The detailed analysis presented in this work reveals the intricate soft and hard X-ray characteristics of the magnetar \src. More specifically, this source exhibits significant energy-dependent variation in pulse profile structure, RMS pulsed fraction, and spectro-polarimetric characteristics, as well as phase-dependent variability in its spectra and polarization. Attempting to parse the system's geometry and emission mechanisms through spectroscopy is nontrivial as the phenomenological spectral components are complexly entangled with each other (see Section \ref{section3}). Examining the spectral properties together with the pulse profile morphology and polarization yields more robust interpretations of the system's soft thermal, soft non-thermal, and hard X-ray emission characteristics, and in turn, a more insightful view of the geometric constraints and radiation mechanisms driving the \src\ broadband emission properties.

\subsection{Soft Thermal Emission}\label{sec:thermal_disc}
The persistently bright, thermal-like soft X-ray emission observed in several magnetars, including \src, can be attributed to the decay of ultra-strong crustal (or toroidal) magnetic field (\citealt{thompsonDuncan_1996}; \citealt{vigano2013MNRAS}). The surface heat distribution is anisotropic, with increased levels of X-ray emission concentrated in regions of higher magnetic energy density from strained surface locales by the decaying toroidal field \citep{2016PNAS..113.3944G}, or return current from twisted, localized B-field bundles \citep{2009ApJ...703.1044B}. Either model is consistent with the quasi-thermal shape of the magnetar spectra, and can qualitatively explain the varying intensity pulse profiles which is mainly dependent on the system geometry \citep{Albano-2010-ApJ,Bernardini-2011-MNRAS}. The double-peaked morphology in \src\ observed in the 0.5--3 keV possibly indicates thermal emission from two quasi-antipodal spots, and the simple simulations presented in Figure~\ref{fig:MTS-fit-J1708} supports this conjecture.

The detailed energy-dependent pulse morphology and phase-resolved spectroscopy reveal more details about the heat topology in this source. The constancy of the RMS pulsed fraction between 0.5--3 keV suggests that the emission in this energy band is dominated by a single mechanism. As revealed by both the phase-averaged and phase-resolved spectroscopy, the most likely origin is the quasi-thermal emission which dominates the soft X-ray band between 0.5--3 keV. The surface temperature experiences a mild evolution as a function of rotational phase ($<20$\%), possibly indicating a mild temperature gradient across the heated surface patch. In contrast, the major change in emitting area with phase (a factor of $\approx5$) and its positive correlation with intensity point to a geometrical effect that is driving the relatively large pulse fraction we observe at soft energies. The pulse profiles subtly evolve at the softest energies, where the broad shoulder at the softest energies (0.5--2 keV) evolves to its own peak at 3 keV; the pulse morphology becomes distinctly double-peaked with peak-separation of 0.5 rotation. This could occur if the emission at these energies is slightly more beamed and/or emanating from a relatively smaller surface area. Such geometrical details, which are not incorporated into the modeling discussed in Section~\ref{sec:modeling}, should provide improved fits to the soft energy pulse profile, and will be treated elsewhere.

\subsection{Soft Nonthermal Emission}\label{sec:SPL_disc}
 The nonthermal emission appearing in magnetar spectra at softer energies ($<10$ keV) is commonly attributed to resonant cyclotron up-scattering (RCS) of thermal photons by mildly relativistic charged electrons/positrons persisting in the closed field lines of a twisted magnetosphere (\citealt{2002ApJ...574..332T}; \citealt{2006RPPh...69.2631H}). This soft nonthermal component modifies the thermal emission before becoming dominant around 3--4 keV. A number of papers have explored various RCS models at the non-relativistic and non-quantum Thomson limit using various simplifying assumptions with general success in reproducing the non-thermal soft X-ray tail (e.g. \citealt{Lyutikov-2006-MNRAS}, \citealt{2007ApJ...660..615F}, \citealt{Nobili-2008-MNRAS}, and \citealt{taverna24Galax} for review). However, RCS depends non-trivially on the structure of the magnetosphere, the  details of which are difficult to constrain. 

The SPL component of \src\ is prominent, dominating the phase-averaged SED from at least 3 keV to 20~keV. Across this same energy range, the pulsed characteristics also change dramatically. Between 4--12~keV, the pulse profile morphology evolves from single-peaked to double-peaked. Meanwhile, the RMS pulsed fraction dips by $\sim 5-10\%$ near 4--5 keV before recovering around 7--8 keV and continuing to increase at 12 keV and beyond. The energy-dependent pulse profiles and spectral analysis together expose the complex interplay of the soft nonthermal emission with the thermal and hard nonthermal components. 

The SPL possesses a distinct primary and secondary peaks which appear more closely associated with the thermal and hard nonthermal emission components, respectively. The DSPs (Figure~\ref{fig:dynamicspectralprofiles}) demonstrate this coupling. The leading SPL peak emerges at $\lesssim2$ keV and largely caps at 12--15 keV, coinciding with disappearance of the initial peak of the energy-dependent pulse profiles. This soft SPL peak lags the BB pulse by $\approx0.1$ cycles and exhibits a quasi-linear phase drift from 1--10 keV of $\Delta\phi\approx0.1$ cycles. These results are consistent with the phase-dependent spectral parameters as evidenced by the component fluxes in the red band of Figure~\ref{fig:spectralparams} and the unfolded $\nu F_{\nu}$ spectra of phases 12--14 and 1--2 in Figure~\ref{fig:phase_resolved_spectra}.

In contrast, the secondary SPL peak decreases in power with energy, yet its phase remains approximately constant. This second SPL peak (or hard SPL peak) manifests at higher energies than the soft SPL peak in the DSP, extending from 2--3 keV up to $\sim 30$ keV, where it overlaps with the hard PL emission. The phase-dependent spectral parameters also show that the hard SPL peak appears aligned with the emission from the hard PL component (Figure \ref{fig:spectralparams}). Moreover, between phases 4--7 of Figure \ref{fig:phase_resolved_spectra}, the spectra appear notably flatter as well.


When considered together, the differences in phase evolution, energy-dependence, and spectral properties of the two SPL peaks suggest that they likely originate from distinct emission regions and possibly different emission mechanisms. The distinct phase-dependent behavior between the two SPL components may reflect differences in the emission geometry, as harder photons are expected to be more strongly beamed at larger loop altitudes, making the observed pulse morphology sensitive to the viewing geometry. The soft SPL peak could arise from reprocessing of the surface emission within the atmosphere or magnetosphere, potentially from magnetospheric resonant cyclotron scattering of the surface thermal photons. Meanwhile, the modeling presented Section \ref{sec:modeling} indicates that the secondary peak associated with harder emission could be produced by particles in higher Landau levels in the upper part of the atmosphere emitting synchrotron radiation.

\subsection{Hard X-ray Tail Emission}\label{sec:HPL_disc}

A number of magnetars, including \src, exhibit non-thermal pulsed hard X-ray tails in the spectra between 20 keV and $\sim 200$ keV that are flatter ($\Gamma \approx 1$) than the non-thermal soft X-ray PL  discussed above. A prevailing mechanism ascribed to the production of these hard X-ray tails is magnetic resonant inverse Compton scattering (RICS) of thermal soft X-ray photons in the atmosphere by relativistic electrons (\citealt{2007Ap&SS.308..109B},\citealt{2011ApJ...733...61B} \citealt{beloborodov13ApJ}, \citealt{wadiasingh18ApJ},\citealt{2019BAAS...51c.292W}). These electrons are accelerated up to Lorentz factors $\gamma \sim 10^3$ likely by potentials in the closed magnetospheric field lines \citep{2005ApJ...634..565T, beloborodov13ApJ}. In such a regime, QED effects, namely the resonant cross-section and kinematics, are necessary to accurately model the hard X-ray ($\gtrsim10$~keV) tail. These scattering interactions most likely occur in the inner magnetosphere, and the resulting emergent hard X-ray spectra will be strongly subjected to the geometry of the system. In highly relativistic RICS, photon splitting and pair production processes become relevant and contribute to the total radiation spectrum \citep{Harding2025ApJ}.


\src\ exhibits a hard PL tail that emerges at approximately $10$ keV and dominates above 20 keV (Figure \ref{fig:phase_averaged_spectra}). At this energy, most of the initial soft peak that was introduced at lower energies in Figure \ref{fig:nuxmm_energy_dependent_pp} has diminished. The duty cycle of the remnant pulse observed from 20 keV to 70 keV notably shrinks by $\sim 15\%$, possibly because at higher energies the HPL component becomes increasingly disentangled from the soft PL component. The DSP in Figure \ref{fig:dynamicspectralprofiles} supports this suggestion; the hard SPL peak overlaps with the HPL flux until the hard SPL peak nearly diminishes entirely at $\approx 30$ keV, whereafter the HPL flux prevails. The DSP elucidates the predominance of the HPL flux output, affirming the suggestion that the majority of the magnetar's bolometric luminosity is emitted at these high energies (\citealt{2005ApJ...634..565T}, \citealt{hartog08AA:0142}). Upon examining the energy-resolved RMS PF in Figure \ref{fig:rms_pf_energy_resolved}, a break occurs around 20 keV where the pulsed fraction continues to increase but at a flatter rate. It may be that the pulsed fraction continues to increase at a less steep ascent all the way to 100\% pulsed fraction as suggested by \citep{2005ApJ...634..565T,2008A&A...489..263D,wadiasingh18ApJ,2022HEAD...1911047W}. When modeled together (Figure \ref{fig:cascade_spectra}), the hard X-ray emission and high polarization near $\approx30$--50 keV is most naturally explained by synchrotron radiation from first-generation pair production (see \ref{sec:modeling}).  However, detailed studies of the full characteristics of the hard X-ray tail requires observations of the suspected spectral peak between 0.1--1 MeV, necessitating an MeV explorer such as COSI \citep{Tomsick2023Xz}, or spectropolarimetry with a larger instrument \citep{2019BAAS...51c.292W}.
    
        
\subsection{Polarization}\label{sec:polarization}

The spectrally decomposed polarization characteristics also reveal a $90$\degree~shift in polarization angle between the BB ($30$\degree) and SPL ($-60$\degree) components. 
The observed PA swing may arise as an artifact of our simple spectro-polarimetric modeling, which assumes energy-independent polarization properties for the BB and SPL components. In the 4–8 keV range, the SPL component is fully dominant in the spectrum (Figure~\ref{fig:phase_averaged_spectra}) and is highly polarized. At lower energies, where the SPL component still contributes substantially but the measured PD is smaller, the model compensates by assigning the BB component an approximately orthogonal polarization angle. This reduces the net polarization in the 2–4 keV band and allows the fit to reproduce the model-independent results \citep[][]{2023ApJ...944L..27Z}. This behavior highlights the limitations of phenomenological spectro-polarimetric models commonly adopted in magnetar analyses, and underscores the need for physically motivated modeling to interpret the polarization results more robustly.



Figure \ref{fig:PA_PD_phases} displays the phase- and energy-resolved polarization characteristics. For reference, the vertical color bars used in Figure \ref{fig:spectralparams} overlay the \ixpe\ profiles, providing a window into the polarization behavior of the spectral components. At low energies, (2--3 and 3--4 keV), the intensity peaks align with the phase interval dominated by the BB emission (depicted by the red band). The PD strongly anti-correlate with the intensity pulse profile, likely caused by the mixing of X-mode and O-mode photons along magnetic field lines, thus causing a net depolarization (Figure~\ref{fig:MTS-fit-J1708}, see also \citealt{heyl24MNRAS:2259}). The PD, while exhibiting a strong phase-dependence, consistently increases according to energy, from a peak of $\approx30$\% between 2--3 keV up to $\approx65$\% in the 4--8 keV band. These values largely agree with the energy-resolved PD discussed above. Intriguingly, the increase in typical PD values corresponds also to the emergence of the shoulder within the phase interval dominated by the HPL emission (blue band). This suggests the component associated with hard emission is likely intrinsically highly polarized.

The modeling presented in Section \ref{sec:modeling} suggests that synchrotron emission from pair-produced plasma can account for \src's high polarization and spectral properties above $\sim4$ keV (Figure \ref{fig:cascade_spectra}). Because single-photon splitting has no energy threshold, synchrotron emission can dominate below $\sim600$ keV only if a restricted set of splitting modes operates. Pair production requires photons to reach energies of $\approx1$ MeV; if all kinematically allowed splitting modes are active, photons may split before reaching this threshold, suppressing pair creation and the resulting synchrotron emission. For the surface magnetic field and viewing geometry adopted in Figure \ref{fig:cascade_spectra}, allowing all three modes instead produces a steeper photon-splitting component that dominates above $\sim50$ keV. This balance depends strongly on geometry and field strength: at smaller observer angles ($30^\circ$--$40^\circ$), emission originates at higher altitudes where the magnetic field is weaker and splitting is less efficient, whereas stronger surface fields allow splitting to dominate over a broader range of viewing angles. Thus, even within \ixpe's 4--8 keV band, polarization measurements can provide an indirect view of the particle kinematics and QED processes that shape magnetar spectra at much higher energies. \src\ is the second magnetar, after 1E 1841--045 \citep{stewart2025ApJ:1841, rigoselli2025ApJ:1841}, for which \ixpe\ polarization measurements have constrained photon-splitting mode kinematics. Additional \ixpe\ observations and future high-energy spectropolarimetry above 10 keV will enable more direct tests of this long-sought QED effect.

\section{Conclusion}
Combining deep broadband X-ray observations with high-energy polarimetry studies offers a new window into magnetar science. The detailed phase-resolved analysis of \src\ enabled by the deep \nustar\ and \xmm\ observations illuminate the complex interplay of the thermal and non-thermal emission exhibited by the source. The X-ray polarization measurements afforded by \ixpe, with phase-resolved polarization degrees ranging between 10\% and 40\%, provide a new avenue to disentangle these spectral signatures and enable fresh studies and insights into the processes underlying their emission.  These results also inform science agendas of future spectro-polarimeters such as COSI and eXTP \citep{Zhang2019:eXTP}. Extending polarimetric observations to larger samples and higher energies will allow us to decisively test magnetar emission models from both their surfaces and inner magnetospheres, and fully exploit these sources as laboratories for extreme physics.

\section{Acknowledgments}

\begin{acknowledgments}
The material is based upon work supported by NASA under award number 80GSFC24M0006. GY acknowledges support through NASA grants 80NSSC25K0283, 80NSSC25K0109, 80NSSC25K0110, and 80NSSC25K7257.
M.G.B. thanks NASA for generous support under awards 80NSSC24K0589, 80NSSC25K7257 and 80NSSC25K0079.  
This work was supported in part by the Big-Data Private-Cloud Research Cyberinfrastructure MRI-award funded by NSF under grant CNS-1338099 and by Rice University's Center for Research Computing (CRC). 
This research is based on observations obtained with XMM-Newton, an ESA science mission with instruments and contributions directly funded by ESA Member States and NASA. 
This research has made use of data and/or software provided by the High Energy Astrophysics Science Archive Research Center (HEASARC), which is a service of the Astrophysics Science Division at NASA/GSFC.  
The authors used ChatGPT (OpenAI) for limited editorial assistance in improving clarity of the text. 
We thank Roberto Taverna and Roberto Turolla for helpful discussions and feedback that informed the analysis and interpretation of the polarization results presented in this work.
\end{acknowledgments}

%

\vspace{5mm}
\facilities{NuSTAR, NICER, XMM-Newton (EPIC-PN), IXPE}


\software{\texttt{astropy} \citep{2013A&A...558A..33A}, IXPEOBSSIM \citep{baldini_2022_softX_ixpe}, NuSTARDAS, NICERDAS, \texttt{nustar-gen-utils} (DOI: 10.5281/zenodo.14969199), Xspec \citep{1996ASPC..101...17A}}



\appendix

\begin{sidewaystable}[h]

\renewcommand{\arraystretch}{1.5}
\centering
\resizebox{1.\textwidth}{!}{
\begin{tabular}{c c c c c c c c c c}
\midrule
Phase & kT & R$^{2*}$ & F$_{\rm BB}^{**}$ & $\Gamma_{\rm soft}$ & F$_{\Gamma_{\rm soft}}^{**}$ & $\Gamma_{\rm hard}$ & F$_{\Gamma_{\rm hard}}^{**}$ & F$_{\rm 0.5-10\  keV}$ & F$_{\rm 10-70\ keV}$ \\
\midrule
Phase-Averaged  & $0.468 \pm 0.003$ & $8.2\pm0.4$ & $(2.81 \pm 0.04)\times 10^{-11}$ & $2.63 \pm 0.04$ & $(9.66 \pm 0.08)\times 10^{-11}$ & $0.5\pm0.1$ & $(1.95 \pm 0.08)\times 10^{-11}$ & $(1.17 \pm 0.008)\times 10^{-10}$ & $(3.04 \pm 0.07)\times 10^{-11}$ \\
Phase 1 & $0.439 \pm 0.006$ & $19.2 \pm 2.0$ & $(4.9 \pm 0.3)\times 10^{-12}$ & $2.59 \pm 0.03$ & $(13.5 \pm 0.5)\times 10^{-12}$ & $0.4^{+0.7}_{-0.8}$ & $(0.9 \pm 0.3)\times 10^{-12}$ & $(13.9 \pm 0.3)\times 10^{-12}$ & $(2.5 \pm 0.2)\times 10^{-12}$ \\
Phase 2 & $0.450 \pm 0.007$ & $13.2 \pm 1.0$ & $(3.8 \pm 0.2)\times 10^{-12}$ & $2.50 \pm 0.03$ & $(11.6 \pm 0.4)\times 10^{-12}$ & $0.1^{+0.5}_{-0.4}$ & $(1.6 \pm 0.3)\times 10^{-12}$ & $(11.4 \pm 0.3)\times 10^{-12}$ & $(3.3 \pm 0.3)\times 10^{-12}$ \\
Phase 3 & $0.44 \pm 0.01$ & $9.4^{+2.0}_{-1.0}$ & $(2.4 \pm 0.2)\times 10^{-12}$ & $2.54 \pm 0.04$ & $(10.1^{+0.4}_{-0.5})\times 10^{-12}$ & $0.9 \pm 0.3$ & $(2.2 \pm 0.3)\times 10^{-12}$ & $(9.4 \pm 0.2)\times 10^{-12}$ & $(3.5 \pm 0.2)\times 10^{-12}$ \\
Phase 4 & $0.45 \pm 0.02$ & $3.97 \pm 1.00$ & $(1.2 \pm 0.2)\times 10^{-12}$ & $2.46^{+0.03}_{-0.04}$ & $(10.1 \pm 0.4)\times 10^{-12}$ & $0.7 \pm 0.3$ & $(3.2 \pm 0.3)\times 10^{-12}$ & $(8.2 \pm 0.2)\times 10^{-12}$ & $(4.6 \pm 0.2)\times 10^{-12}$ \\
Phase 5 & $0.42 \pm 0.02$ & $4.7 \pm 1.0$ & $(1.1 \pm 0.2)\times 10^{-12}$ & $2.25 \pm 0.03$ & $(10.1^{+0.2}_{-0.3})\times 10^{-12}$ & $0.3 \pm 0.4$ & $(2.8 \pm 0.4)\times 10^{-12}$ & $(7.4 \pm 0.2)\times 10^{-12}$ & $(5.2 \pm 0.3)\times 10^{-12}$ \\
Phase 6 & $0.47 \pm 0.02$ & $3.6^{+0.9}_{-0.8}$ & $(1.2 \pm 0.2)\times 10^{-12}$ & $2.34 \pm 0.04$ & $(8.9^{+0.3}_{-0.4})\times 10^{-12}$ & $0.6 \pm 0.2$ & $(4.2 \pm 0.4)\times 10^{-12}$ & $(7.2 \pm 0.2)\times 10^{-12}$ & $(5.9 \pm 0.3)\times 10^{-12}$ \\
Phase 7 & $0.442 \pm 0.009$ & $9.0 \pm 1.0$ & $(2.4 \pm 0.2)\times 10^{-12}$ & $2.46 \pm 0.04$ & $(7.3 \pm 0.3)\times 10^{-12}$ & $0.4 \pm 0.3$ & $(3.3 \pm 0.3)\times 10^{-12}$ & $(7.3 \pm 0.2)\times 10^{-12}$ & $(4.4 \pm 0.2)\times 10^{-12}$ \\
Phase 8 & $0.47 \pm 0.01$ & $6.3^{+1.0}_{-0.9}$ & $(2.1 \pm 0.2)\times 10^{-12}$ & $2.88 \pm 0.04$ & $(7.5 \pm 0.4)\times 10^{-12}$ & $0.8 \pm 0.2$ & $(2.6 \pm 0.2)\times 10^{-12}$ & $(7.4 \pm 0.2)\times 10^{-12}$ & $(2.8 \pm 0.2)\times 10^{-12}$ \\
Phase 9 & $0.438^{+0.008}_{-0.007}$ & $11.6 \pm 1.0$ & $(3.0 \pm 0.2)\times 10^{-12}$ & $2.70 \pm 0.04$ & $(7.4 \pm 0.4)\times 10^{-12}$ & $-0.3 \pm 0.4$ & $(1.4 \pm 0.3)\times 10^{-12}$ & $(7.9^{+0.3}_{-0.2})\times 10^{-12}$ & $(2.1^{+0.3}_{-0.2})\times 10^{-12}$ \\
Phase 10 & $0.482^{+0.010}_{-0.009}$ & $6.7^{+0.9}_{-0.8}$ & $(2.5 \pm 0.2)\times 10^{-12}$ & $2.76 \pm 0.04$ & $(8.0 \pm 0.4)\times 10^{-12}$ & $0.4^{+0.3}_{-0.4}$ & $(1.4^{+0.3}_{-0.2})\times 10^{-12}$ & $(8.0 \pm 0.2)\times 10^{-12}$ & $(2.0 \pm 0.2)\times 10^{-12}$ \\
Phase 11 & $0.483 \pm 0.007$ & $9.0^{+0.9}_{-0.8}$ & $(3.4 \pm 0.2)\times 10^{-12}$ & $2.74 \pm 0.03$ & $(8.6 \pm 0.4)\times 10^{-12}$ & $-0.2 \pm 0.5$ & $(1.2 \pm 0.3)\times 10^{-12}$ & $(9.2 \pm 0.3)\times 10^{-12}$ & $(1.9 \pm 0.2)\times 10^{-12}$ \\
Phase 12 & $0.505 \pm 0.006$ & $9.9^{+0.8}_{-0.7}$ & $(4.5 \pm 0.2)\times 10^{-12}$ & $2.73 \pm 0.04$ & $(8.2 \pm 0.4)\times 10^{-12}$ & $0.4 \pm 0.3$ & $(1.8 \pm 0.2)\times 10^{-12}$ & $(10.1 \pm 0.3)\times 10^{-12}$ & $(2.5 \pm 0.2)\times 10^{-12}$ \\
Phase 13 & $0.505 \pm 0.006$ & $10.7 \pm 0.8$ & $(4.9 \pm 0.2)\times 10^{-12}$ & $2.84 \pm 0.03$ & $(9.8 \pm 0.4)\times 10^{-12}$ & $0.6 \pm 0.3$ & $(1.6 \pm 0.2)\times 10^{-12}$ & $(11.5 \pm 0.3)\times 10^{-12}$ & $(2.2 \pm 0.2)\times 10^{-12}$ \\
Phase 14 & $0.501 \pm 0.008$ & $8.6^{+0.9}_{-0.8}$ & $(3.8 \pm 0.2)\times 10^{-12}$ & $2.73^{+0.02}_{-0.03}$ & $(13.1 \pm 0.4)\times 10^{-12}$ & $0.2 \pm 0.5$ & $(1.1^{+0.3}_{-0.2})\times 10^{-12}$ & $(12.7 \pm 0.3)\times 10^{-12}$ & $(2.2 \pm 0.2)\times 10^{-12}$ \\
\midrule
\hline
\end{tabular}}
\begin{list}{}{}
    \item[{\bf Notes.}]$^{*}$Derived by adopting a 3.85 kpc distance. \\ $^{**}$Derived in the 0.5--70 keV range, with units of $\mathrm{10^{-11} \text{erg s}^{-1}\text{ cm}^{-2}}$. \\ 
    All uncertainties represent the 68\% confidence interval. Fluxes are absorption-corrected.  
    \end{list} 
    \caption{Phase-averaged and phase-resolved BB+2PL spectral components and fluxes.}
\label{table:phase_averaged_and_phase_resolved_parameters}

\end{sidewaystable}

\newpage

\begin{table}[h!]
\centering
\begin{tabular}{c c}

\hline
Model Parameter & Value\\ 

\hline
PD$_{\text{BB}}$ & $0.51 \pm 0.08$\\
PA$_{\text{BB}}$ & $31${\degree}$
\pm 4$ \\
kT (keV) & $0.464 \pm 0.002$\\
R$^2$ (km$^2$) & $9.1\pm0.3$ \\
PD$_{\Gamma_{\mathrm{soft}}}$ & $0.84 \pm 0.05$ \\
PA$_{\Gamma_{\mathrm{soft}}}$ & $-61${\degree}$
\pm 1 $\\
$\Gamma_{\mathrm{soft}}$ & $2.62 \pm 0.01$\\
norm$_{\mathrm{soft}}$ & $0.0260 \pm 0.0004$ \\
$\Gamma_{\mathrm{hard}}$ & $0.6 \pm 0.1$\\
norm$_{\mathrm{hard}}$ & $(4^{+2}_{-1}) \times 10^{-5}$ \\
\midrule
\hline
\end{tabular}
\caption{\src\ phase-averaged spectrally decomposed absorbed BB+2PL model with a constant polarization component added to each parameter, i.e. \texttt{const*tbabs(bbodyrad*polconst+pow*polconst+pow)} Xspec model. These parameters were obtained through simultaneously fitting \nustar, \xmm, and \ixpe\ data of \src\ (with fixed N$_{\text{H}}$). The associated errors are obtained through the \texttt{error} command in Xspec at 1$\sigma$.}
\label{table:spectral_decomposition}
\end{table}

\bibliographystyle{aasjournal}



\end{document}